\documentclass[fleqn,usenatbib]{mnras}

\usepackage[T1]{fontenc}

\DeclareRobustCommand{\VAN}[3]{#2}
\let\VANthebibliography\thebibliography
\def\thebibliography{\DeclareRobustCommand{\VAN}[3]{##3}\VANthebibliography}

\usepackage{array,multirow,graphicx} 
\usepackage{float} 
\usepackage{amsmath}  
\usepackage{amssymb}  
\usepackage{physics}
\usepackage{tabularx}
\usepackage{xcolor}
\usepackage{tikz}
\usepackage{tablefootnote}
\definecolor{lime}{HTML}{A6CE39}
\DeclareRobustCommand{\orcidicon}{%
  \begin{tikzpicture}
  \draw[lime, fill=lime] (0,0) 
  circle [radius=0.16] 
  node[white] {{\fontfamily{qag}\selectfont \tiny ID}};
  \draw[white, fill=white] (-0.0625,0.095) 
  circle [radius=0.007];
  \end{tikzpicture}
  \hspace{-2mm}
}

\foreach \x in {A, ..., Z}{%
  \expandafter\xdef\csname orcid\x\endcsname{\noexpand\href{https://orcid.org/\csname orcidauthor\x\endcsname}{\noexpand\orcidicon}}
}

\usepackage{newtxtext,newtxmath}

\newcommand\REPLY[1]{\textcolor{black}{#1}}

\title[Gal-gal strong lensing statistics in surveys]{A model for galaxy-galaxy strong lensing statistics in surveys}

\author[G. Ferrami \& J.S.B. Wyithe]{
G. Ferrami $^{1,}$$^{2}$ \thanks{E-mail: gferrami@student.unimelb.edu.au } \orcidA{}
J. Stuart B. Wyithe $^{1,}$$^{2,}$$^{3}$ \orcidB{}\\
$^{1}$School of Physics, University of Melbourne, Parkville, VIC 3010, Australia\\
$^{2}$ARC Centre of Excellence for All-Sky Astrophysics in 3 Dimensions (ASTRO 3D)\\
$^{3}$Research School of Astronomy and Astrophysics, Australian National University, Canberra, ACT 2611, Australia}

\date{Accepted XXX. Received YYY; in original form ZZZ}

\pubyear{2024}

\begin{document}
\label{firstpage}
\pagerange{\pageref{firstpage}--\pageref{lastpage}}
\maketitle

\begin{abstract}
Photometric wide-area observations in the next decade will be capable of detecting a large number of galaxy-scale strong gravitational lenses, increasing the gravitational lens sample size by orders of magnitude. 
To aid in forecasting and analysis of these surveys, we construct a flexible model based on observed distributions for the lens and source properties and test it on the results of past lens searches, including SL2S, SuGOHI and searches on the COSMOS HST and DES fields.
We use this model to estimate the expected yields of some current and planned surveys, including Euclid Wide, Vera Rubin LSST, and Roman High Latitude Wide Area.
The model proposed includes a set of free parameters to constrain on the identifiability of a lens in an image, allowing construction of prior probability distributions for different lens detection methods.
The code used in this work is made publicly available.
\end{abstract}

\begin{keywords}
gravitational lensing: strong -- galaxies: high-redshift -- galaxies: evolution
\end{keywords}



\section{Introduction}

Strong gravitational lensing is a powerful probe of both astrophysics and cosmology, as its effects depend on the surface mass density of the lens and the cosmological distances between the observer, lens and source.
Strong lensing has been used to study the inner structure of galaxies (e.g., \citealt{Zwicky_lenses}, \citealt{Treu_Koopmans_2002}, \citealt{Koopmans_Treu_2003}) 
and their evolution with cosmic time (e.g., \citealt{Grillo_SLACS_2009}, \citealt{Sonnenfeld_SL2S_2013}, \citealt{Beyond_bulge_halo}, \citealt{Project_Dinos}), 
as well as to constrain the dark matter fraction in massive ETGs (e.g., \citealt{Grillo_DM_fraction}, \citealt{Sonnenfeld_DM_fraction_SL2S}), 
their stellar initial mass function (IMF, e.g., \citealt{Treu_IMF}, \citealt{SINFONI_survey}, \citealt{Sonnenfeld_IMF_SuGOHI}) 
and the presence of substructures within them and along their line-of-sight (e.g., \citealt{Substructure_lens_galaxy}, \citealt{Vegetti_Substructure_Lensing}, \citealt{Oldham_DM_fraction}, \citealt{Despali_substructure}). 
Galaxy scale strong lenses have been used to study the high redshift quasar luminosity function (e.g., \citealt{Turner_1984}, \citealt{Webster_AGN_LF}, \citealt{Wyithe_Loeb_AGN_LF}) and galaxy luminosity function, especially its bright end (e.g., \citealt{Wyithe_Nature_2011}, \citealt{Barone_Nugent_2015}, \citealt{Mason_2015}, \citealt{Ferrami_Lensing_bright_end}).
Lensing can also be a tool to measure the values of cosmological parameters independently from the distance ladder, from either observation of time-delay between multiple images of the same source (e.g., \citealt{Refsdal_1964}, \citealt{Suyu_COSMOGRAIL}, \citealt{H0LiCOW_Wong}, \citealt{Birrer_TDCOSMO}, \citealt{Shajib_lensing_H0}), joint strong lensing and dynamical analysis (e.g., \citealt{Grillo_cosmo_param_lens_dyn}), lenses with multiple source planes (e.g., \citealt{Jackpot_Gavazzi}, \citealt{Collett_Auger_DoubleSrc}), or from source and deflector populations (e.g., \citealt{Oguri_2012}).
On kiloparsec scales, strong gravitational lensing, in combination with stellar kinematics of the lens, is sensitive to the weak-field metric of gravity and can be used as a test for General Relativity (e.g., \citealt{Schwab_Bolton_Rappaport}, \citealt{Collett_GR_test}).
For many of these analyses, the small sample size dominates the statistical uncertainty.
A large sample of lenses with sources at different redshifts is critical to marginalise over any trends in the redshift evolution of the deflector population and would unlock a broader range of investigations accessible with greater statistical power.
However, galaxy scale lenses are difficult to observe, as they require alignment within $\sim 1$" of a light source and a foreground object with a sufficiently large surface mass density, along with deep and high-resolution observations of a large fraction of the sky to identify rare cases.

Since the first strong lens was discovered (Q 0957+561; \citealt{Walsh_first_observed_lens}), a combination of serendipitous discoveries and systematic searches made the number of known galaxy-scale lenses steadily increase: 11 in the early 1990s (\citealt{Blandford_Narayan_1992}), around 200 in 2010 (\citealt{Treu_review_2010}), and $\mathcal{O}(10^3)$ galaxy-scale lenses found to date\footnote{The exact number depends on the cut in purity applied to the sample, as many of these lenses have high probability of being lenses, but no spectroscopic confirmation.} (e.g., \citealt{Sonnenfeld_citizien_science}).
During the next decade, $\mathcal{O}(10^5)$ strong lenses (\citealt{Collett_2015}, \citealt{Marshall_Blandford_Sako}, \citealt{Holloway_2023}) will be discovered by wide-field photometric surveys offering improved depth, area, and resolution compared to existing data.
These include the Euclid Wide survey (\citealt{Euclid_Wide_Survey}), Vera Rubin Observatory LSST (\citealt{Vera_Rubin_LSST}), and Roman Space Telescope High Latitude Wide Area Survey (\citealt{Roman_WFIRST}). 

Spectroscopic follow-up of a candidate lens is necessary for two reasons: first to distinguish morphological features that appear similar to strong lenses (e.g., ring galaxies, star forming tidal streams) from genuine galaxy scale lenses, and second to convert the angular quantities of the putative lenses into physical units.
The 4MOST Strong Lensing Spectroscopic Legacy Survey (4SLSLS, \citealt{4SLSLS_COLLETT}) will provide spectroscopic redshifts
for $\mathcal{O}(10^4)$ lens-sources pairs, and velocity dispersion measurements for $\sim 5000$ lenses.\\
\indent Such large samples of strong lenses will unlock considerable scientific potential through vastly improved statistics 
(e.g. \citealt{Sonnenfeld_SL_Stat_I_innerstructure}; \citealt{Sonnenfeld_SL_Stat_III_inference_complete}, \citealt{Sonnenfeld_SL_Stat_IV_inference_incomplete}). 
To tackle the forthcoming thousand-fold increase in data volume, model inference must be automated, and made robust without human intervention (e.g., \textsc{PyAutoLens} \citealt{Nightingale_Dye_PyAutolens}, \citealt{No_lens_left_behind}; or \textsc{Dolphin} \citealt{Shajib_Dolphin}).\\
\indent Many methods have been used to find galaxy scale strong lenses, mostly from photometric datasets.
These techniques include feature detection algorithms (e.g., \textsc{arcfinder}, \citealt{ARCFINDER_Alard}, \citealt{ARCFINDER_More};  \textsc{ringfinder}, \citealt{RINGFINDER_ALGORITHM}; or \textsc{bna}, \citealt{Diehl_DES_search}, \citealt{DES_Strong_lens_odonnel});
lens model fitting algorithms (e.g. \citealt{Marshall_model_detection}, \citealt{Chan_model_detection});
a combination of the previous two (e.g., \textsc{yattalens}, \citealt{Sonnenfeld_SuGOHI_I}); 
principal component analysis to subtract galaxies from imaging data (e.g., \citealt{PCA_lens_search}); 
visual search by researchers (\citealt{Hogg_lens_in_HST}, \citealt{Moustakas_visual_inspection}, \citealt{Faure_2008}, \citealt{Pawase_visual_inspection}); citizen science (\citealt{Marshall_citizien_science}; \citealt{More_citizien_science};  \citealt{Sonnenfeld_citizien_science}).
Recently, machine learning has been commonly used as a new tool.
Some of the most common architectures employed so far in lens searches are Convolutional Neural Networks (CNNs, see for example \citealt{Jacobs_CNN_CFHTLS}, \citealt{Jacobs_CNN_DES}, \citealt{He_CNN_search}, \citealt{CNN_unspervised_search}, \citealt{Rosati_CNN_search}, \citealt{EUCLID_CNN_lensfinder}), Support Vector Machines (SVM, e.g., \citealt{Support_Vector_Machines}) and self-attention encoding (e.g., \citealt{Self_attention_ML}).
Note that the techniques listed above can be combined to classify a lens sample (e.g., an ensemble classifier based on neural network and citizen science lens finders in \citealt{Holloway_ensemble_classifier}).\\
\indent Given a certain intrinsic distribution in redshift and mass of lenses, any given search measures the distribution of lensed sources \textit{identifiable} given the observational constraints which could be then biased by the telescope in use and the adopted search method.
In general, it is non-trivial to infer the intrinsic distribution of lenses given the observed one (e.g., \citealt{Sonnenfeld_SL_selection}).
Past efforts to characterize the intrinsic distribution of {identifiable} lenses for a given survey include analytical models for lensing statistics of bright galaxies in spectroscopic searches (\citealt{Serjeant_analytical_model}) and AGNs and supernovae in photometric surveys (\citealt{Oguri_Marshall}), and simulation-based models for lensing statistics of galaxy-galaxy lensing (\citealt{Collett_2015}, \citealt{Holloway_2023}).\\
\indent In this paper we present a flexible analytic model to forecast the expected yields of some current and planned surveys using fiducial distributions for the lens and source properties as constrained by observations.
The model proposed includes a set of free parameters to explicitly set the constraints on the identifiability of a lens in an image, considering both the cases where the lens light can and cannot subtracted. 
This paper is organized as follows.
In Section 2, we introduce the model for galaxy scale strong lens statistics. 
In Section 3, we consider the response of our model to variations in its input parameters.
In Section 4, we compare our model results to some past strong lens search samples. 
In Section 5, we present forecasts for the number of strong lenses in ongoing and future surveys.
In Section 6, we estimate the fraction of rare quad configurations and dual-plane lenses.
In Section 7, we discuss potential improvements to the lens and source population models, and their effect on extreme lens configurations (i.e., sources at high redshift and/or highly magnified).
Conclusions are presented in Section 8.
Throughout this paper, we adopt $H_0 = 70$ km s$^{-1}$ Mpc$^{-1}$, $\Omega_0 = 0.3$,  $\Omega_\Lambda= 0.7$.

\section{A model for galaxy-galaxy lenses statistics}

In this paper, we present a model for the statistics of galaxy-galaxy lenses.
Our model considers only information contained in a single photometric band.
We first calculate the total number of lensed galaxies in a given patch of sky.
Then we construct a flexible tool to estimate the number of \textit{identifiable} lensed sources.
Initially, we assume that the lens light can be completely subtracted.
We then relax this condition to account for the effect of lens light.

\subsection{Lensing probability and Magnification Bias}

The \textit{a-priori} probability for a source to be lensed by a foreground galaxy into multiple images is defined as the multiple image optical depth $\tau$ (\citealt{Turner_1984}), calculated as the integral of the lensing cross section over the lens population and redshift.
Under the usual approximation that the population of lenses is dominated by early-type galaxies (e.g., \citealt{Turner_1984}, \citealt{Kochanek_1996}, \citealt{Oguri_Marshall}), we take the mass distribution profile of the lenses to be a Singular Isothermal Ellipsoids (SIE), which has been shown to be a very reliable approximation for the total matter surface density profile at the radial scales relevant for galaxy-galaxy lensing (\citealt{Gavazzi_SLACS_2007}, \citealt{Koopmans_bulge_halo_conspiracy}, \citealt{Lapi_2012}, \citealt{Sonnenfeld_SL2S_2013}).

The strong lensing magnification effects of SIE lenses averaged over the possible source positions, depend only on the velocity dispersion $\sigma$ of the stellar component (a proxy for the total mass) and on the ellipticity of the SIE mass distribution. 
We can calculate $\tau$ for a population of sources at redshift $z_s$ given by the lenses intervening on the line-of-sight at each redshift $z_l<z_s$ as
\begin{equation}\label{eq:optical_depth}
\begin{split}
  \tau (z_s)= C\int_0^{z_s} \dd z_l\int_0^\infty \dd \sigma \Phi(\sigma,z_l)(1+z_l)^3
  \frac{c \dd t}{\dd z_l} \pi D_l^2 \theta_E^2(\sigma,z_l) \text{ ;}
\end{split}
\end{equation}
where $C$ is a lens model dependent constant ($C=1$ for a singular isothermal sphere, $C<1$ for a singular isothermal ellipsoid lens), 
$\Phi(\sigma,z_l)$ is the velocity dispersion function (VDF) of early-type galaxies, $D_l$ is the angular diameter distance of the lens and $\theta_E$ is its Einstein radius.
We adopt the \cite{Mason_2015} velocity dispersion function and its evolution with redshift $\Phi(\sigma,z_l)$ with their best-fit values for the relevant parameters.
In the next Sections, we will explore two other VDFs: the local Velocity Dispersion Function measured by SDSS (\citealt{Choi_2007_SDSS}), and the distribution calibrated on a sample of strong lenses (\citealt{Geng_2021}).
We included the dependency on the ellipticity distribution of lenses using the geometry constraints up to $z=2$ provided by the SDSS (\citealt{van_der_Wel_2014}), following the same method used in \cite{Ferrami_Lensing_bright_end}.
Averaging over the observed early-type galaxies ellipticity distribution, we find $C\approx0.9$.

The luminosity function (LF) describes the comoving density of galaxies with magnitude between $M$ and $M+\dd M$.
The LF is often approximated by a Schechter profile
\begin{equation}\label{eq:Schechter_LF}
  \Psi(L)\dd L = \Psi_\star \left(\frac{L}{L_\star}\right)^\alpha\exp\left(-\frac{L}{L_\star}\right)\frac{\dd L}{L_\star} \text{ ,}
\end{equation}
and the parameters $\Psi_\star, \alpha, L_\star$ evolve with redshift.
We adopt the rest frame UV LF evolution presented in \cite{Bouwens21_data} for $z\leq9$.
Strong lensing can produce an apparent magnification of the source, which in turn alters the observed luminosity function of high redshift sources (e.g., \citealt{Turner_1984}, \citealt{1995_Pei}, \citealt{Wyithe_Nature_2011}).
The lensing bias $B$ quantifies the excess of lensed sources of magnitude $M$ at redshift $z_s$ is
\begin{equation}\label{eq:Lensing_Bias}
  B(M, zs) = \frac{1}{\Psi(M, z_s)} \int_0^\infty \frac{\dd \mu}{\mu}\dv{P}{\mu}\Psi\left(M+\frac{5}{2}\log(\mu), z_s\right)  \text{ .}
\end{equation}
Here $\Psi(M, z_s)$ is the unlensed galaxy luminosity function and $\dv{P}{\mu}$ represents the magnification distribution of the brightest image produced by a lens with isothermal mass distribution.
Following the finding of \cite{Ferrami_Lensing_bright_end} that source size only affects magnification bias for the brightest galaxies, in this paper we do not account for the effect of source size in the magnification bias, since the main contribution to the observed lens numbers comes from its faint end. 
The difference between the observed flux in a given bandpass and the rest frame magnitude given by the UV LF is determined by the galaxy spectral energy distribution (SED) and the bandpass profile.
The apparent luminosity of a galaxy at redshift $z$ observed in a bandpass $b$ is related to absolute UV magnitude as
\begin{equation}\label{eq:Apparent_magnitude}
  m_b = M_{UV} + DM(z) + K_{UV,b}(z) \:\text{,} 
\end{equation}
where $DM(z)$ is the cosmological distance modulus and $K_{UV,b}(z)$ is the K-correction.

In this work, we approximate the K-correction from the UV continuum slope $\beta$ ($f_\lambda \propto \lambda^\beta$) of a star-forming galaxy SED. 
Calibrating the SED continuum slope from the observations accounts for the effects of dust attenuation in the galaxy rest frame.
For sources with redshift in the range $0<z_s<4$ we adopt the linear fit $\beta = -1.5-0.12(z_s-1)$ over the magnitude range $-19<M_{UV}<-20$ shown in \cite{UV_cont_near_un}.
For the source redshift range $4 \leq z \leq 10$, we adopt the piece-wise linear relations between $\beta$ and $M_{UV}$ studied \cite{Bouwens_UV_cont_slopes} (see also Sect. 2.3 of \citealt{Dragons_IV_UVLF} for a detailed implementation of these relations).

\subsection{Total lensed background}

For a lens with fixed velocity dispersion $\sigma$ and redshift $z_l$, the number density of background sources with absolute magnitude $M$ at redshift $z_s$ it lenses is obtained as the product between the luminosity function and the comoving volume behind this single lens (enclosed in the portion of the source plane that can produce multiple images).
In a flat universe, this equates to
\begin{equation}\label{eq:Prob_lens_backgnd}
  \frac{\dd N_{SL}(M, z_s | \sigma, z_l)}{\dd z_s \dd M}= C\pi  \theta_E^2 D_c^2(z_s)\frac{c}{H(z_s)} \Psi(M, z_s) B(M, zs) \:,
\end{equation}
where $D_c$ is the comoving distance to the source.

Assuming a survey with area $A_s$ in square degrees, integration over the lens population and redshift and the source magnitude and redshift gives the total number of lensed sources in a given field,
\begin{equation}\label{eq:Number_lens_galaxies}\begin{aligned}
  N = & 
  \int_0^{z_{s,max}} \dd{z_s} 
  \int_0^{M_\text{cut}(z_s)} \dd{M} 
  \int_0^{z_s} \dd{z_l} 
  \int_0^\infty \dd{\sigma} 
  \dv{V(A_s, z_l)}{z_l}\\
  &\times \frac{\dd N_{SL}(M, z_s | \sigma, z_l)}{\dd z_s \dd M}
  \Phi(\sigma, z_l) \:,
\end{aligned}\end{equation}
where $z_{s,max}$ is the maximum redshift considered for the source population, $M_\text{cut}(z_s)$ is the magnitude cut\footnote{
In general, $m_\text{cut}$ is chosen to be fainter than the survey magnitude limit $m_\text{lim}$ to obtain higher completeness in the sample.}
applied the survey data $m_\text{cut}$ corrected for the distance modulus, K-correction, and $\dv{V}{z_l}$ is the section of the comoving volume shell behind $A_s$.

Note that by construction Eq. \ref{eq:Prob_lens_backgnd} accounts only for the region behind a lens that produces multiple images (i.e., a magnification $\gtrsim 2$ for a SIE lens).

\subsection{Detecting arcs and counterimages}

In lens searches the identification of a lens can also be through features, such as extended arcs where a second image is not detected, or two or more images with the same colour detected around $\approx 1"$ around the candidate lens.
In this section, we derive an analytical model that accounts for the minimum magnification of the brightest image (i.e. minimum tangential stretch of an arc) and the apparent luminosity of a counterimage.

\begin{figure}
  \includegraphics[width=\linewidth]{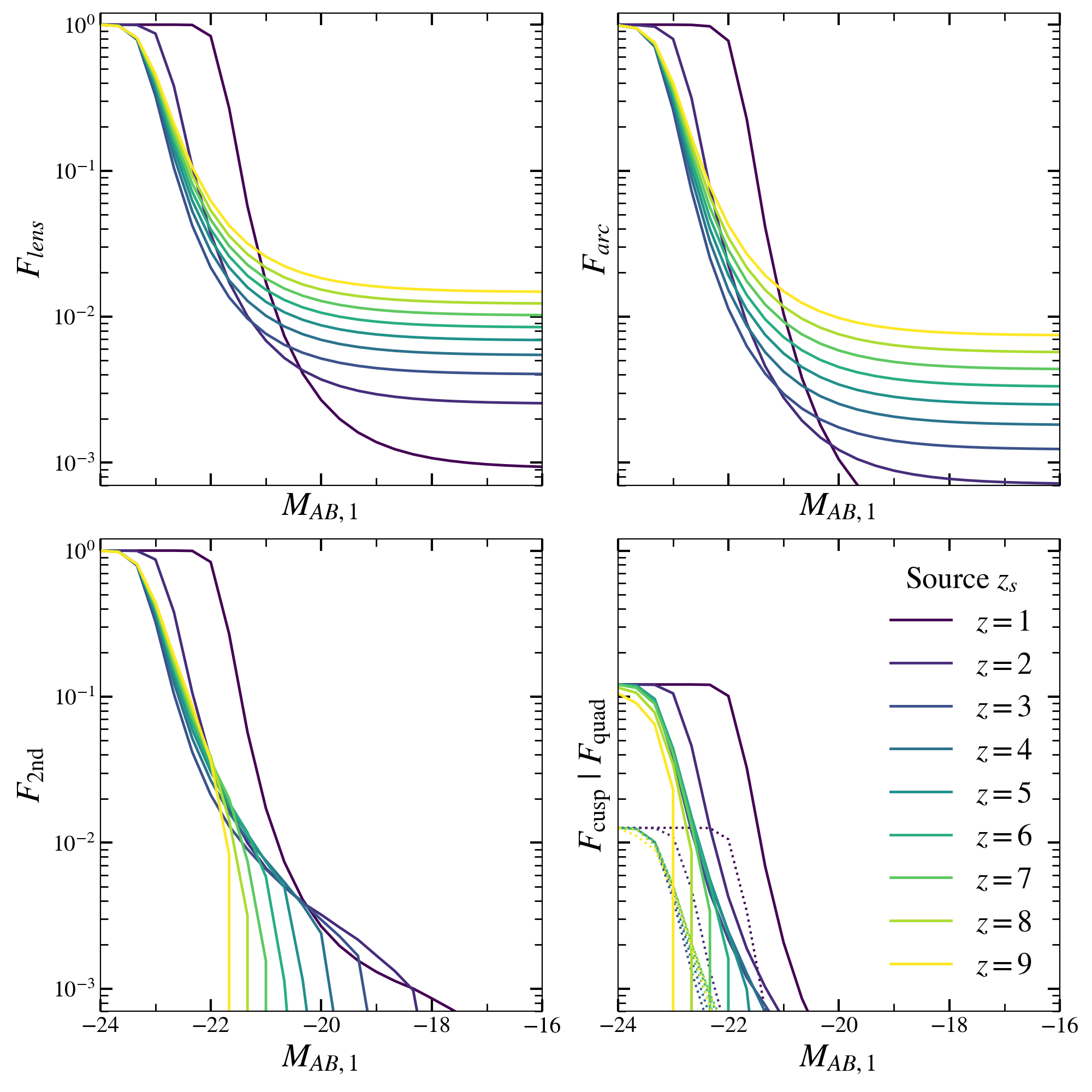}
    \caption{Fraction of lensed objects (top left) as a function of source redshift and absolute UV magnitude. 
    The top left panel shows the fraction of multiply imaged sources (see Eq. \ref{eq:Lens_Fraction}), while the top right panel contains the profile for the fraction of sources imaged into an arc stretched by at least a factor of 3 (Eq. \ref{eq:Arc_Fraction}). 
    The bottom panels show the fraction of second images (left) and third/fourth images in a cusp/quad configuration (right) above an apparent limit magnitude of $m_\text{lim}=28.5$.}
    \label{fig:F_lens}
\end{figure}

Assuming $B=1$ outside $\tau$, the lensed fraction for a given source magnitude and redshift is 
\begin{equation}\label{eq:Lens_Fraction}
  F_\text{lensed}(M, zs) = \frac{\tau B(M, zs)}{\tau B(M, zs)+(1-\tau)} \:.
\end{equation}
The fraction of lenses with a magnification greater than some value $\mu_a$ is then
\begin{equation}\label{eq:Arc_Fraction}
  \frac{F_\text{arc} (\mu>\mu_a)}{F_\text{lensed}} = 
  \frac{\int_{\mu_a}^\infty \frac{\dd{\mu}}{\mu}\dv{P}{\mu}\Psi\left(M+\frac{5}{2}\log(\mu), z_s\right)}
  {\int_{\mu_\text{min}}^{\infty} \frac{\dd{\mu}}{\mu}\dv{P}{\mu}\Psi\left(M+\frac{5}{2}\log(\mu), z_s\right)} \:,
\end{equation}
where $\mu_\text{min}$ is the lowest possible magnification for the brightest image in the multiple image regime.
Similarly, the fraction of lenses with the luminosity of the $n$-th image of a SIE lens above the magnitude limit is
\begin{equation}\label{eq:Nth_Fraction}
  \frac{F_{n\text{th}} (M_{n}>M_\text{lim})}{F_\text{lensed}} = 
  \frac{\int_{\mu_{n,L}}^{\mu_{n,U}} \frac{\dd{\mu}}{\mu}\dv{P}{\mu} A_n\Psi\left(M+\frac{5}{2}\log(\mu), z_s\right)}
  {\int_{\mu_\text{min}}^{\infty} \frac{\dd{\mu}}{\mu}\dv{P}{\mu}\Psi\left(M+\frac{5}{2}\log(\mu), z_s\right)} \:,
\end{equation}
where $\mu_{n,L}$ and $\mu_{n,U}$ are respectively the lower and upper magnification limit for the brightest image that corresponds to a magnification of the $n$-th image high enough to be brighter than the survey's flux limit.
The fraction $F_n$ is properly normalised by $A_n$, which is the portion of the area within the outermost caustic that can produce a $n$-th image.
In a SIS lens, there is an analytical relation between the magnification of the first and second images ($\mu_2 = \mu_1 - 2$) and therefore getting $F_\text{2nd}^{SIS}$ is straightforward.
Since such an analytical solution is not available for a generic SIE model, we compute an effective relation between the magnification of the primary image and the magnification of the second, third (in the naked cusp regime), and fourth image.
To do so, we first compute the magnification distribution functions for the four images of a SIE as a function of the lens ellipticity.
Then we map those functions into each other through an integral relation ($\int_{-\infty}^{\mu_1} \dv{P}{\mu_1}\dd{\mu} = \int_{-\infty}^{\mu_{n}} \dv{P}{\mu_{n}}\dd{\mu}$) and use that to implicitly solve for the equation $\log(\mu_n/\mu_1) = 0.4(M_1-M_\text{lim})$.
In Figure \ref{fig:F_lens} we plot such lens fraction statistics as a function of the magnitude of the primary image and the source redshift.

Accounting for the identification of arcs and counterimages introduces a correction to Eq. \ref{eq:Prob_lens_backgnd}. 
We specifically introduce a factor of $\mathcal{F}_\text{arc} = F_\text{arc}/F_\text{lensed}$ and $\mathcal{F}_{n\text{th}} = F_{n\text{th}}/F_\text{lensed}$ as so far we were considering all the sources within the outer caustic.

\subsection{Selection on minimum SNR and angular size}

The possibility of detecting a galaxy also depends on the ability to collect enough of its photons to overcome the noise in the detector.
This imposes a minimum Signal-to-Noise Ratio (SNR) that constrains the number of \textit{identifiable} lenses. 
In the same way, the angular size of the lensed images must be resolved in order to identify a lens.
This requires the Einstein radius to be larger than the seeing or PSF of a given survey.

\subsubsection{Signal-to-noise ratio}
For a given detector, the number of photons from a source of apparent magnitude $m_b$ is
\begin{equation}\label{eq:n_photons_mag}
N_\gamma = t_\text{exp} 10^{\left(\frac{ZP-m_b}{2.5}\right)}\:,
\end{equation}
where the zero point $ZP$ is the magnitude of an object that produces one count per second and $t_\text{exp}$ is the total exposure time of the survey.
Since we do not have a set of pixels to evaluate the individual contribution to the SNR, we approximate it over the apparent size of the source in the sky as
\begin{equation}\label{eq:SNR}
\text{SNR} = \frac{N_\gamma}{\sqrt{N_\gamma + \sigma_\text{sky}^2}} \:,
\end{equation}
where $\sigma_\text{sky}^2 = N_\text{sky}\pi \vartheta_e^2$ is the survey background noise.
$N_\text{sky}$ is the background photon count per steradian, obtained from the sky background magnitude $m_\text{sky}$ as in Eq. \ref{eq:n_photons_mag}, and the effective radius in radians $\vartheta_e$ comes from the luminosity-size relation, calculated on the intrinsic UV magnitude. 
We use the $M_{UV}-r_e$ relation evolution with redshift presented in \cite{Shibuya_L_size_rel}:
\begin{equation}\label{eq:Lum_size_relation}
  r_e = r_0 \left(\frac{L_{UV}}{L_0}\right)^\alpha \:, \: \left[\alpha = 0.27,\: r_0 = 6.9(1+z)^{-1.2}\:\text{kpc}\right]
\end{equation}
where $r_0$ value represents the effective radius expressed in kpc at a luminosity of $L_0$ corresponding to $M_{UV} = -21$.
We can then evaluate the $\text{SNR}_1$ ($\text{SNR}_2$) for the brightest (secondary) image in a lens knowing that its average magnification in a SIS lens is $\bar\mu_1=3$ ($\bar\mu_2 = 1$).
We can now define two selection functions $\mathcal{S}_{1/2} = H(\text{SNR}_{1/2} > \text{SNR}_\text{min})$, where $H$ is the Heaviside step function, imposing a cut-off for the lenses that have the first or second image below a minimum value of SNR.

\subsubsection{Einstein radius vs. seeing / PSF}
In a lens galaxy, the multiple images appear at an angular scale of the order of the Einstein radius $\Theta_E$. 
For a lens to be identifiable, the background images must be resolved, i.e. the Einstein radius must be bigger than the seeing (or PSF) of the observations.
We define a Heaviside selection function $\mathcal{S}_E = H(\theta_E > \xi s)$, excluding the lenses with an Einstein radius smaller than $\xi$ times the seeing $s$.

\subsection{Number of lenses with lens light subtracted}

We can use the expressions for the fraction of lenses with a clear arc and/or with the second brightest image above the flux limit derived in Eqs. \ref{eq:Arc_Fraction}-\ref{eq:Nth_Fraction} 
as a weighting to evaluate the number of the \textit{detectable} lensed sources in the background of a given lens,
\begin{equation}\label{eq:Prob_lens_backgnd_with_fractions12}
\begin{aligned}
\frac{\dd N_{SL}^{obs}}{\dd z_s \dd M}(M, z_s, \mu_a, M_\text{lim}) =
\frac{\dd N_{SL}}{\dd z_s \dd M} 
\times \max(\mathcal{S}_1\mathcal{F}_\text{arc}, \mathcal{S}_2\mathcal{F}_\text{2nd}) \mathcal{S}_E \:,
\end{aligned}
\end{equation}
where the selection functions $\mathcal{S}_{1/2}$ and $\mathcal{S}_E$ have been defined in the previous Section.
This constraint should give the bulk of the identifiable lenses, even though for ground-based surveys (i.e. large seeing) it does not retrieve the (small) portion of lenses that can be resolved after deconvolving the PSF.
The maximum $\max(F_\text{arc}, F_\text{2nd})$ approximates the detection selection up to a factor of order $2$ in the luminosity interval where $F_\text{arc} \approx F_\text{2nd}$.

Substituting $N_{SL}^{obs}$ to $N_{SL}$ in Eq. \ref{eq:Number_lens_galaxies} gives the number of lensed sources at redshift $z_s$ with magnitude $M$ detectable by a given survey, assuming that the lens light can be removed without loss of information.

\subsection{Accounting for the lens light profile}

In previous sections, we calculated the number of gal-gal lenses in a survey in which the lens light is assumed to be completely removed so that the limiting factors in the observability of a lensed image are flux limit and seeing of a specific survey.

We now consider lens galaxies in which the lens light has not been subtracted.
Such lens systems may prove difficult to identify, especially for bright lenses (favoured in the lens statistics as they often are massive, i.e. good deflectors) and for faint sources (e.g., \citealt{Kochanek_1996}; \citealt{Wyithe_SDSS_2002}). 
We can introduce the effect of the lens light profile in our model by comparing the surface brightness (SB) of the source (a conserved quantity in lensing) to the SB of the lens galaxy, evaluated at the position of the multiple images.

To do so, we will use the Fundamental Plane relation for early-type galaxies (\citealt{Djorgovski_Davis_1987}, \citealt{Dressler_FP_1987}, see also \citealt{2001TreuFP}):
\begin{equation}\label{eq:Fundamental_Plane}
  \log R_e = \alpha \log_{10}\sigma_0 + \beta \langle\mu_e\rangle + \gamma \text{ ,}
\end{equation}
where $R_e$ is the effective radius, $\sigma_0$ is the central velocity dispersion and $\langle\mu_e\rangle$ is the mean surface brightness measured within the effective radius.
The FP parameters in the \textit{grizYJHK} wavebands are taken from \cite{BarberaFPgrizYJHK}.
The evolution with lens redshift of the three parameters is given by a linear interpolation between redshift $z=0$ and $z=2$ in the B-band (\citealt{Stockmann_2021}, see Table 2). We further assume that the FP parameters evolve with the same slope in every rest-frame wavelength. 

To account for the lens light profile, we obtain the surface brightness sampling the Fundamental Plane for a given $\sigma_0 = \sigma$ and a log-gaussian distribution of effective radii with mean and variance that depends on the lens rest frame emission waveband as described in \cite{LaBarbera_SPIDER_I_Reff} (see in particular Fig. 11 of that paper), with the mean evolving with redshift following the same trend as the scale-radius in the $M_{UV}-R_e$ relation from \cite{Shibuya_L_size_rel} (see Sect. 2.4).

The parameters of the FP are taken from the lens rest-frame waveband that would be redshifted to observing waveband $b$, to be specified for each survey.
We then consider a deVaucouleur profile (\citealt{DeVaucouleurs1948}, \citealt{DeVaucouleurs1953}) for the SB of the lens and evaluate it at the position of the bright and secondary images, and compare it with the SB of the source (accounting for cosmological dimming).
To obtain a simple analytical form for the position of the bright and secondary images, we model the lens mass density as a Singular Isothermal Sphere (SIS). 
In a SIS lens, the distance between the first and second image is always $\Delta \theta = 1 \theta_E$, and the position of the brightest image is uniformly distributed between $\theta_E < \theta_1 < 2\theta_E$.

We calculate the surface brightness of a source of magnitude $m_b$ using the $M_{UV}-R_e$ relation from \cite{Shibuya_L_size_rel} introduced in Sect. 2.4.
We then include the probabilities of seeing the first and second image through the lens light, $P_1^{LL}$ and $P_2^{LL}$, into Eq. \ref{eq:Prob_lens_backgnd_with_fractions12} as
\begin{equation}\label{eq:Prob_lens_backgnd_with_fractions_LL}\begin{aligned}
  \frac{\dd N_{SL}^{obs, LL}}{\dd z_s \dd M}&(M, z_s, \mu_a, M_\text{lim}, b) =\\
  &\frac{\dd N_{SL}}{\dd z_s \dd M} \times 
  \max(\mathcal{S}_1\mathcal{F}_\text{arc}P_1^{LL}, \mathcal{S}_2\mathcal{F}_\text{2nd}P_2^{LL})\mathcal{S}_E \text{ .}
\end{aligned}\end{equation}
Substituting $N_{SL}^{obs, LL}$ to $N_{SL}$ in Eq. \ref{eq:Number_lens_galaxies} gives the number of lensed sources at redshift $z_s$ with magnitude $M$ detectable by a given survey without removing the lens light.

This approach provides a lower bound on the number of observable lenses, as colour information from observations in multiple photometric bands of the same field can yield higher fractions of detected lenses than this method based on the SB alone. 
\footnote{
  For background galaxies hosting an active AGN or a supernova event, luminosity time variation accessible by cadenced surveys can also increase the fraction of identifiable lensed sources.}
On the other hand, the model without lens light introduced in the previous section provides an upper bound on the number of identifiable lenses.
The combination of these two estimates could provide some insight into the response of the completeness function on the survey characteristics.

\begin{table}
  \centering
  \caption{Summary of parameters and functions that constitute the input of our model. 
  When values are indicated, these are used throughout the paper. 
  Otherwise, the values indicated with a dash are survey-dependent (they can be found in Tables \ref{tab:Survey_ZP_N_sky} - \ref{tab:Survey_results}).
  The value of $\mu_\text{arc}$, $\xi$, and SNR$_\text{min}$ are chosen to match the lens identifiability prescriptions adopted in \citealt{Collett_2015} and \citealt{Holloway_2023}.}
  \label{tab:Summary_param}
  \begin{tabular*}{\linewidth}{@{\extracolsep{\fill}} l c l }
    \hline 
    Parameter        & Value   & Description                                       \\
    \hline 
    \hline 
    $\mu_\text{arc}$ & 3       & Minimum magnification of bright image             \\
    $\xi$            & 1.5     & Seeing threshold factor                           \\
    SNR$_\text{min}$ & 20      & Minimum SNR for lens detection                    \\
    $m_\text{lim}$   & -       & 5$\sigma$ magnitude limit of the survey           \\
    $m_\text{cut}$   & -       & Cut in magnitude  \\
    $A_\text{S}$     & -       & Survey area in square degrees                     \\
    $s$              & -       & Seeing in arcsec                                  \\
    $b$              & -       & Observing photometric band                        \\
    $ZP$             & -       & Survey zero point in the band $b$                 \\
    $t_\text{exp}$   & -       & Total exposure time                               \\
    $m_\text{sky}$   & -       & Sky background magnitude                          \\
    \\
    \hline 
    Functions        & Ref     &                                                   \\
    \hline 
    \hline 
    SIE              & \scriptsize${1}$& Lens total mass profile\\
    $\Psi(L, z)\dd L$& \scriptsize${2}$& Source UV Luminosity function\\
    $M_{UV}$-$R_e$   & \scriptsize${3}$& Luminosity size relation\\
    $\Phi(\sigma, z)$& \scriptsize${4, 5}$& Velocity dispersion function \\
    Ell. Distr.      & \scriptsize${6}$& Lens ellipticity distribution\\
    FP               & \scriptsize${7}$& Fundamental plane parameters \\
    \hline 
    \end{tabular*}
  \begin{flushleft}
  \footnotesize \textit{Notes.} $^1$\cite{Kormann_Schneider_Bartelmann_1994}, $^{2}$\cite{Bouwens21_data}, $^{3}$\cite{Shibuya_L_size_rel},
  \footnotesize $^4$\cite{Mason_2015}, $^{5}$\cite{Geng_2021}, $^{6}$\cite{van_der_Wel_2014}, $^{7}${\cite{BarberaFPgrizYJHK}}
  \end{flushleft}
\end{table}

\begin{figure*}
  \includegraphics[width=\linewidth]{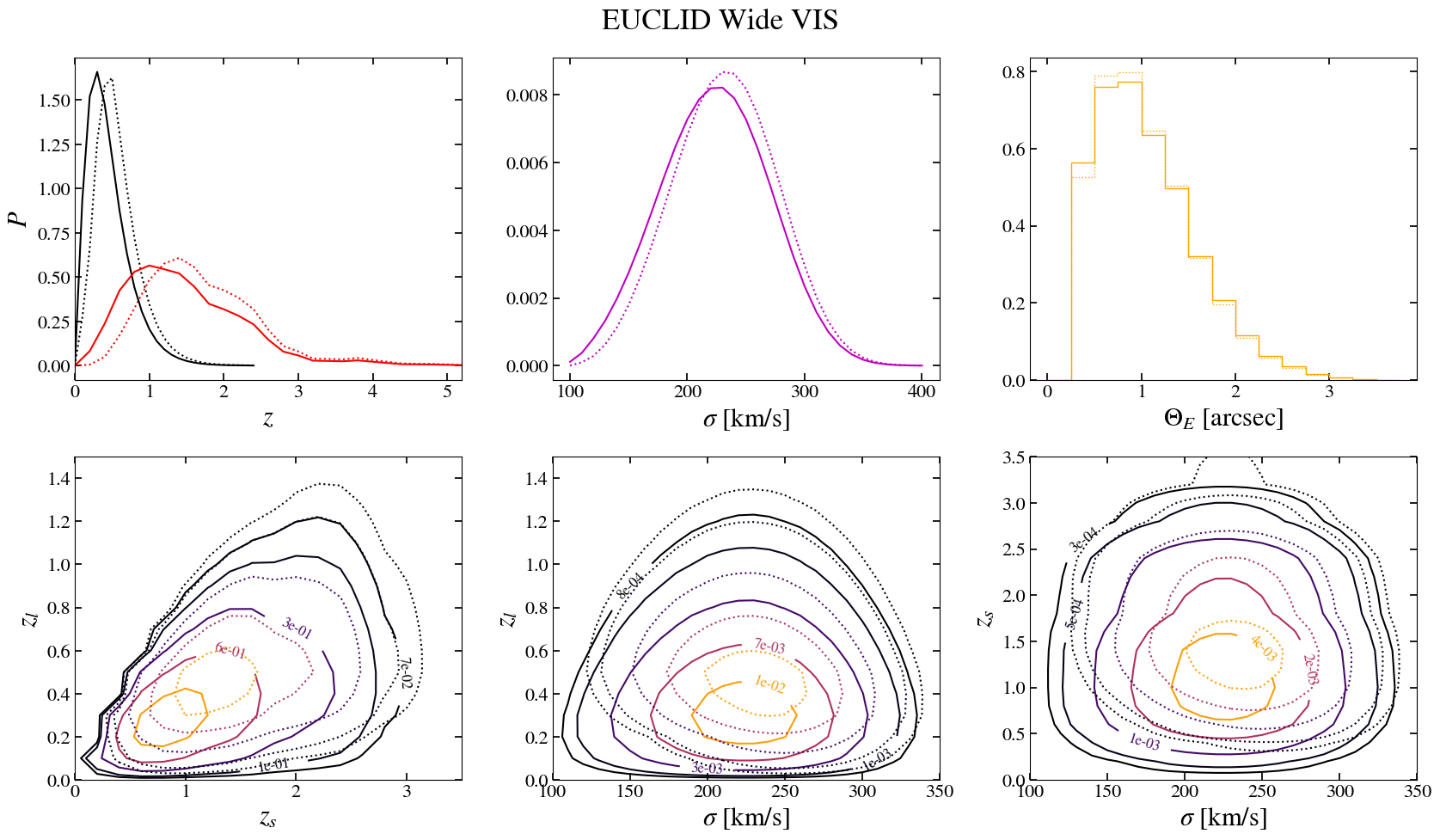}
    \caption{Statistical properties of the lens-source populations our model predicts for EUCLID Wide VIS. The plot shows the distributions assuming lens light can be completely removed (solid lines) and accounting for lens light (dotted lines).
    The top left panel shows the redshift distributions of the lens (black) and source (red) populations. The middle and right upper panels show the velocity dispersion and the Einstein radius distributions, respectively.
    The bottom row panels show how the $z_l$, $z_s$ and $\sigma$ distributions are linked, representing their pairwise \REPLY{probability} contour plots.
    }
    \label{fig:EUCLID_stat}
\end{figure*}

Table \ref{tab:Summary_param} summarizes the input parameters and distributions that enter in our model. 
For an example of the output of the model based on the Euclid Wide VIS band, see Fig. \ref{fig:EUCLID_stat} \REPLY{(a discussion about the forecast for future surveys, Euclid included, can be found in Section 5 and the results are summarised in Table \ref{tab:Survey_results})}. 
The figure shows the projections of the lens number density over the parameter space of redshifts and velocity dispersion ($z_l$, $z_s$ and $\sigma$).

\begin{table}
  \caption{ Survey zero-points used for the lens frequency estimates and value for the uniform sky background. 
  For the Roman telescope, we use the same value inferred from the filter transmission curves by \citealt{Holloway_2023}.}
  \label{tab:Survey_ZP_N_sky}
  \centering
  \begin{tabular*}{\linewidth}{@{\extracolsep{\fill}} l l l l c }
      \hline 
      Telescope  & Filter   & ZP [mag] & m$_\text{sky}$ & Ref \\
      \hline 
      \hline 
      HST        & F814W    & 25.95    & 20             & \scriptsize$1$\\
      \hline 
                 & F115W    & 27.59    & 22             &               \\
      JWST       & F150W    & 27.89    & 22             & \scriptsize$2$ and \scriptsize$3$\\
                 & F277W    & 27.98    & 22             &               \\ 
      \hline 
      CHFT       & i        & 26.24    & 19.2           & \scriptsize$4$\\
      \hline
      DECam      & i        & 30       & 20.1           & \scriptsize$4$\\
      \hline
      SUBARU HSC & i        & 30       & 19.2           & \scriptsize$4$\\
      \hline 
                 & VIS      & 25.50    & 22.2           & \scriptsize$4$\\
      EUCLID     & Y        & 25.04    & 22             &               \\
                 & J        & 25.26    & 22             & \scriptsize$5$\\
                 & H        & 25.21    & 22             &               \\
      \hline
      Roman      & J129     & 26.40    & 22             & \scriptsize$6$\\
      \hline
      Vera Rubin & i        & 30       & 20.1           & \scriptsize$7$\\
      \hline
  \end{tabular*}
  \begin{flushleft}
  \footnotesize \textit{Notes.} $^1$\url{https://acszeropoints.stsci.edu/}, 
  \footnotesize $^2$\url{https://jwst-docs.stsci.edu/jwst-near-infrared-camera/nircam-performance/nircam-absolute-flux-calibration-and-zeropoints},
  \footnotesize $^3$\cite{Windhorst_2022_PEARLS},
  \footnotesize $^4$same values used in \cite{Collett_2015},
  \footnotesize $^5$\cite{Euclid_Wide_Survey}, 
  \footnotesize $^6$\citealt{Roman_WFIRST},
  \footnotesize $^7$\citealt{LSST_Ivezic_2019}.
  \end{flushleft} 
\end{table}

\section{Model response to input parameters}

The main advantage of an analytic model over a simulation-based approach is the ability to vary all the input parameters and fiducial distributions at little computational cost.
In this section, we discuss how different assumptions on the lens population and some of the model input parameters listed in Table \ref{tab:Summary_param} influence the predicted lens and source distribution in mass and redshift, as well as the total number of lenses forecast by the model.

\begin{figure*}
  \includegraphics[width=\linewidth]{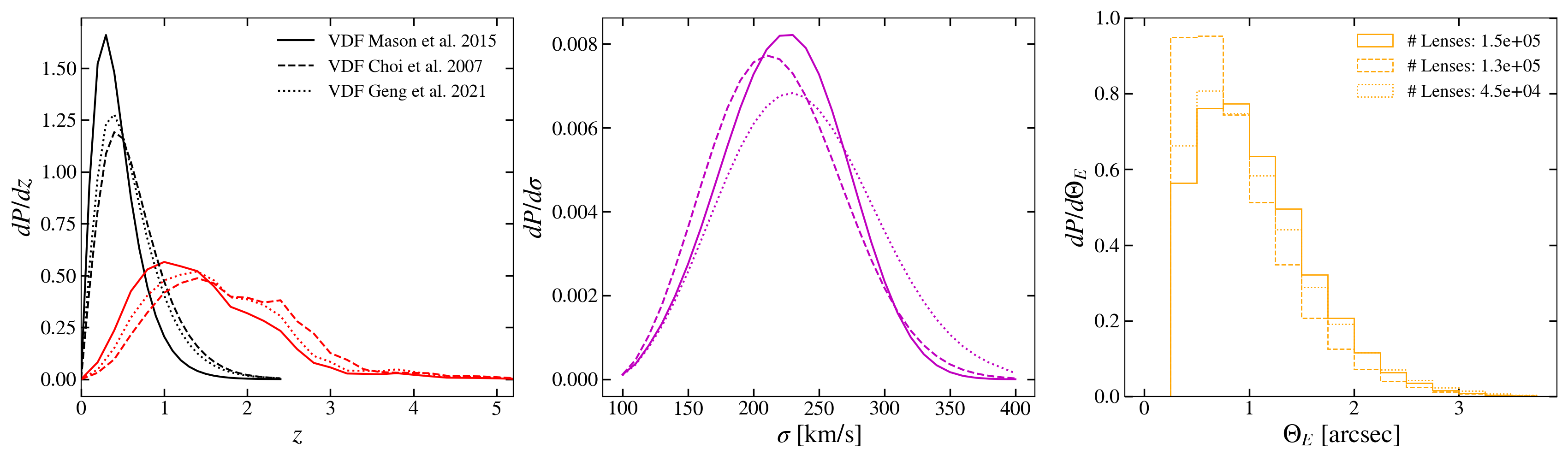}
    \caption{Properties of the lens population as a function of the velocity dispersion function (VDF) profile. 
    The left panel shows the distributions of lens (black) and source (red) redshift for three different VDFs: \citealt{Mason_2015} (the fiducial distribution assumed in our model, solid line), \citealt{Choi_2007_SDSS} (dashed), and \citealt{Geng_2021} (dotted).
    The middle panel shows the velocity dispersion distribution. The right panel shows the distribution of Einstein Radii sizes. Here the legend lists the total number of lenses discoverable in Euclid for each VDF considered.}
    \label{fig:effect_VDF}
\end{figure*}

As an illustration Fig. \ref{fig:effect_VDF} shows the predicted distributions of lens system properties for EUCLID using three VDFs: the model by \cite{Mason_2015}, a constant VDF calibrated in the local universe from \cite{Choi_2007_SDSS}, and the VDF inferred from a sample of gravitational lenses up to $z\sim1$ presented in \cite{Geng_2021}.
As expected we see that the lens velocity dispersion and redshift distributions are very sensitive to the choice of VDF, with our fiducial model producing a narrower $\dv{P}{z_l}$ peaking at a lower redshift compared to the other two VDFs considered.
The distribution of Einstein ring sizes changes accordingly. 

\begin{figure*}
  \includegraphics[width=\linewidth]{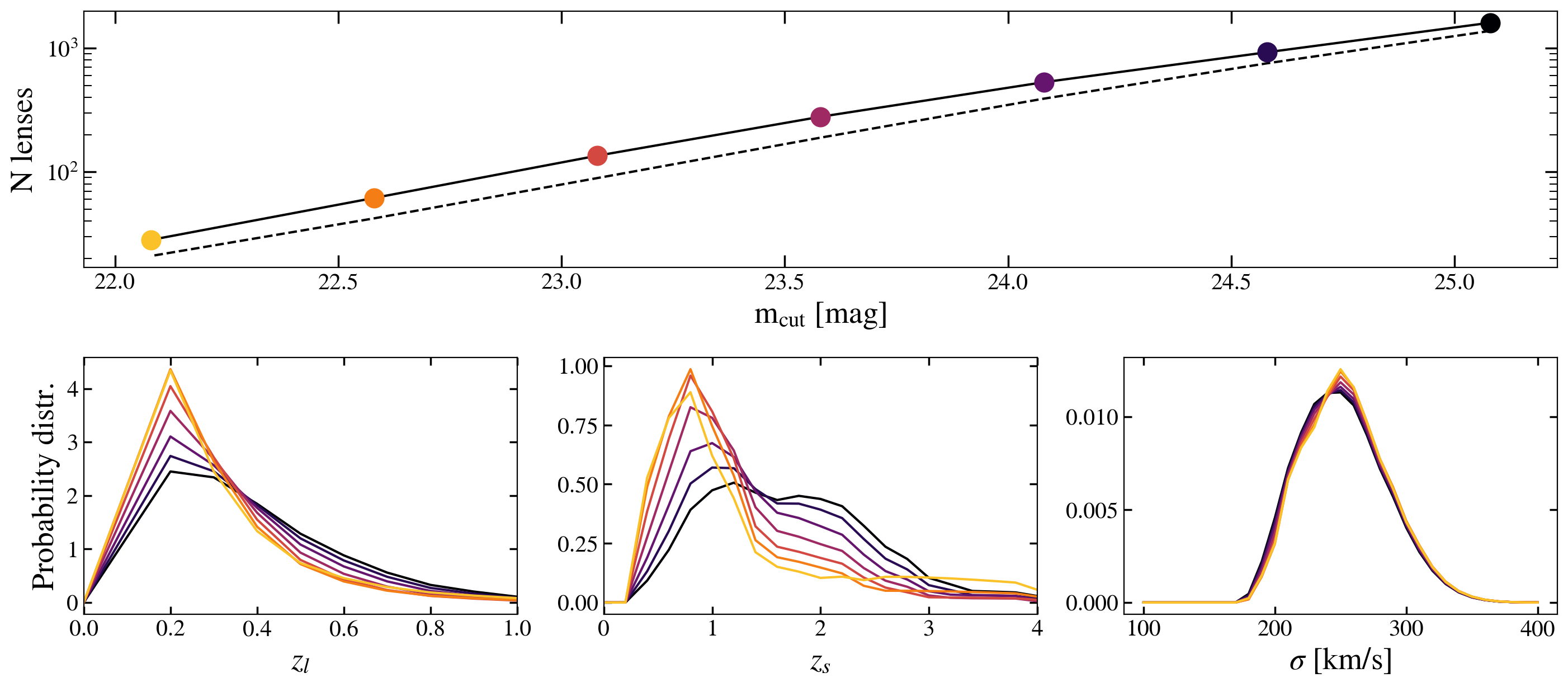}
    \caption{Properties of the lens population as a function of the imposed cut in magnitude. 
    The top panel shows the number of lenses for different choices of $m_\text{cut}$ (coloured dots) assuming that lens light can be removed (solid line) or with lens light in the sample (dashed line). Following the same colour associated with the value of $m_\text{cut}$ in the top panel, in the bottom panels are shown the distributions in lens redshift (left), source redshift (middle) and lens velocity dispersion (right).}
    \label{fig:effect_mcut}
\end{figure*}

We summarize the effect of imposing a magnitude cut in Fig. \ref{fig:effect_mcut}. With decreasing $m_\text{cut}$ (i.e., requiring brighter images), the number of lenses decreases smoothly (the rate at which it decreases changes depending on the survey parameters and value of apparent magnitude), while the source redshift distribution becomes skewed to lower redshift (since the distance modulus shifts more distant sources outside the range of visible magnitudes).
This drives a bias to lower redshifts in the lens population, while not affecting the velocity dispersion distribution. 
\footnote{Changing the limiting magnitude $m_\text{lim}$ would have very similar effects, though it is not completely degenerate with $m_\text{cut}$ as the first enters in the 
observable fraction of second images in Eq. \ref{eq:Nth_Fraction}, while the second gives the integration upper limit over magnitude in Eq. \ref{eq:Number_lens_galaxies}.}

\begin{figure*}
  \includegraphics[width=\linewidth]{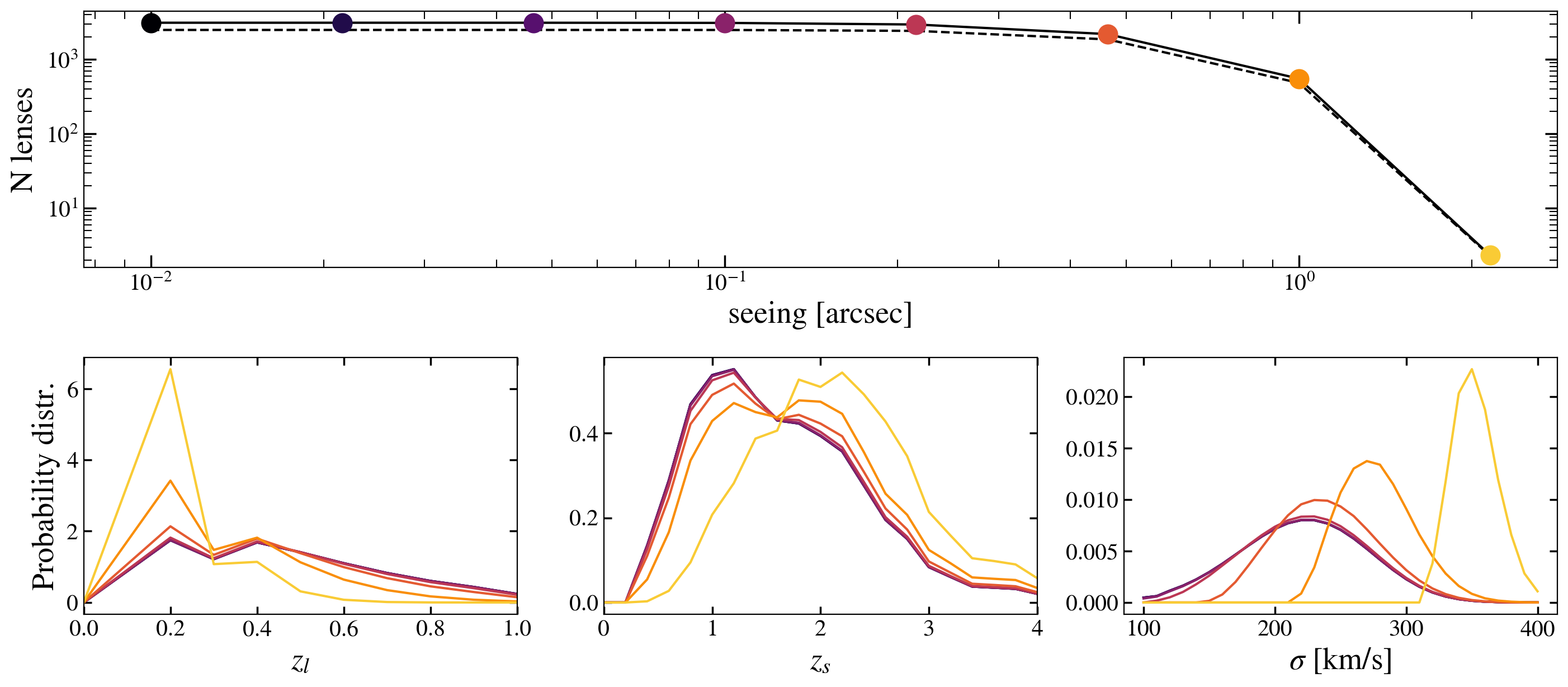}
    \caption{Properties of the lens population as a function of the seeing (or PSF) of the survey. Same panel composition and colour code as in Fig. \ref{fig:effect_mcut}.}
    \label{fig:effect_seeing}
\end{figure*}

\begin{figure*}
  \includegraphics[width=\linewidth]{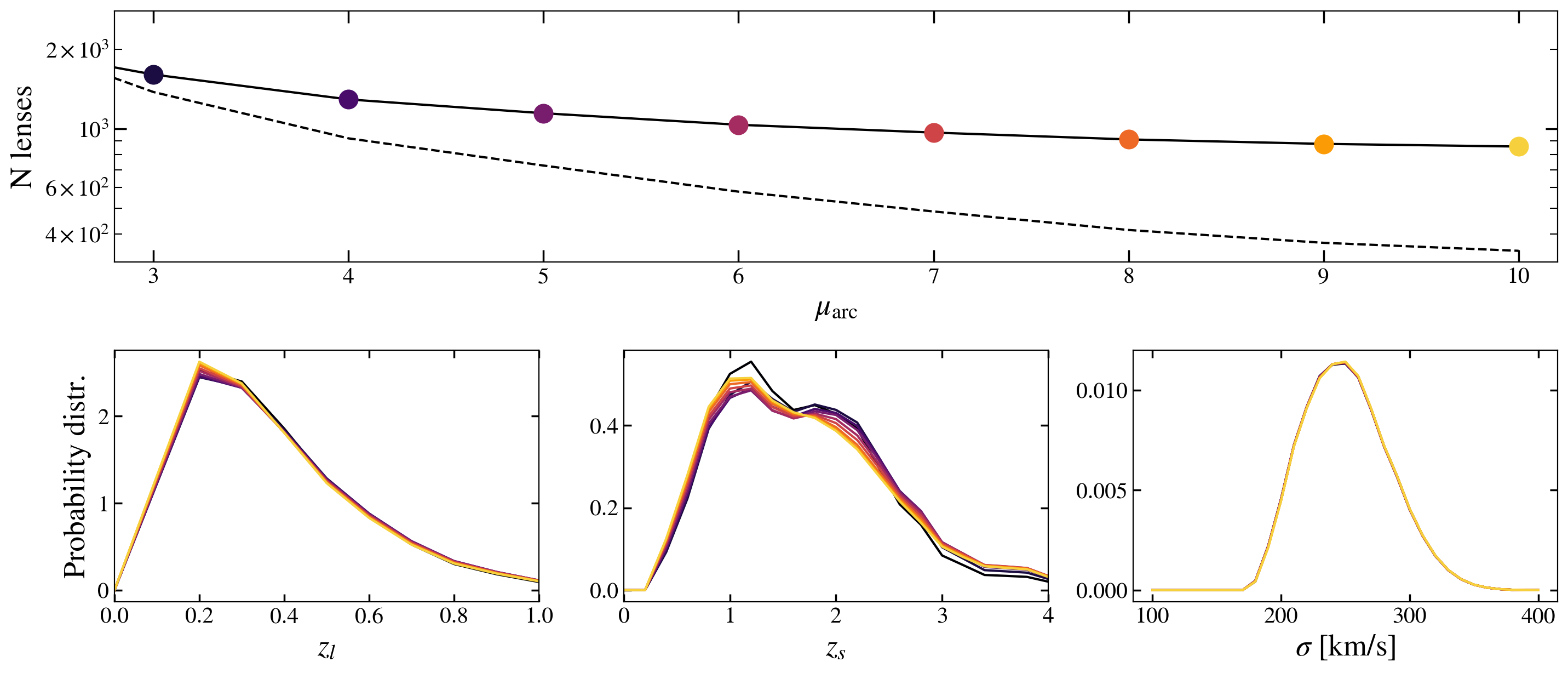}
    \caption{Properties of the lens population as a function of the minimum magnification required for the brightest lensed arc. Same panel composition and colour code as in Fig. \ref{fig:effect_mcut}.}
    \label{fig:effect_muarc}
\end{figure*}

Fig. \ref{fig:effect_seeing} shows that the effect of seeing (or PSF) is negligible as long as it is much smaller than the typical size of the Einstein ring in arcsec. When the two angular quantities become comparable, the survey statistics degrade rapidly, favouring the more massive galaxies (i.e., biasing towards large values of velocity dispersion), and shifting lens and source redshift in opposite directions to maximize the ratio of cosmological distances $D_{ls}/D_{s}$. 
This is completely degenerate with the seeing threshold factor $\xi$.
Finally, increasing the value of the minimum magnitude of the brightest arc necessary to identify a lens decreases the number of discoverable lenses, while mildly biasing the source redshift distribution, as presented in Fig. \ref{fig:effect_muarc}.

\section{Testing against observed lens samples}

In this section, we compare our results to three well-studied lens samples, each obtained with a different lens search technique: the lens catalogue obtained from a targeted visual search in the COSMOS HST field presented in \cite{Faure_2008}; 
the Strong Lensing Legacy Survey (SL2S, \citealt{CFHTLS_SL2S}) using the final sample presented in \cite{Sonnenfeld_SL2S_2013}; the CNN-based search in the DES catalogue from \cite{Jacobs_catalog}; and the Survey of Gravitationally-lensed Objects in HSC Imaging (SuGOHI,  
\citealt{Sonnenfeld_SuGOHI_I}).\\
The summary of the results is listed in Table \ref{tab:Past_Searches_results}.

\subsection{COSMOS HST}
\begin{figure*}
  \includegraphics[width=\linewidth]{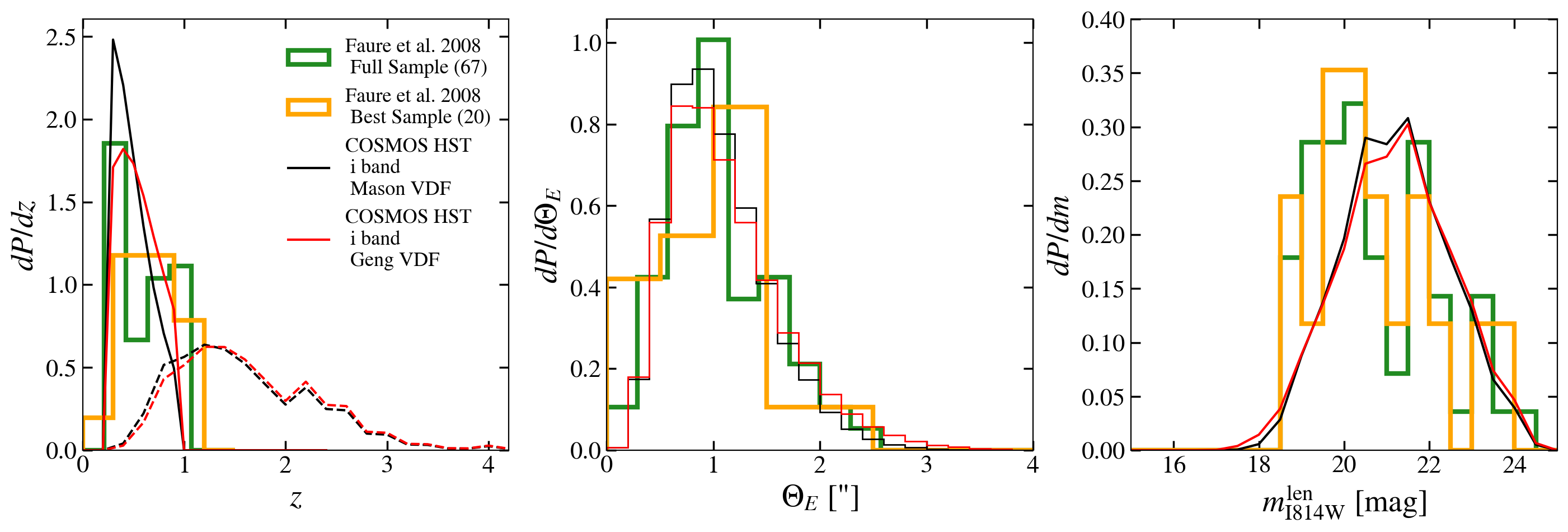}
    \caption{The lens sample statistics from visual inspection on the HST COSMOS data (\citealt{Faure_2008}, full sample in thick green line, best sample in thick orange line) compared to our model expectations for the HST COSMOS $i$ band assuming the VDF from \citealt{Mason_2015} (in black) or \citealt{Geng_2021} (in red).
    The left panel shows the redshift distributions of the lens (solid) and source (dashed) populations. The middle panel shows the distributions of Einstein radii. The right panel shows the distributions of apparent magnitudes of the lens sample.
    The data bin width is based on the Freedman–Diaconis rule.}
    \label{fig:COSMOS_HST_Faure}
\end{figure*}

The Hubble Space Telescope COSMOS survey (\citealt{HST_COSMOS}) observed $1.64$ deg$^2$ of sky in the I814W band.
In Fig. \ref{fig:COSMOS_HST_Faure} we compare the lens statistics predicted by our model for this field to the catalogue of lenses found in COSMOS HST through visual inspection by \cite{Faure_2008}.
After an initial cut in magnitude to maximize completeness ($m_\text{I814W}<25$), their observational catalogue was built in four steps: select potential lenses among bright ($M_V < -20$) foreground ($0.2 < z_\text{phot} < 1$) ETGs, inspecting 10"$\times$10" cutouts around the candidate lenses to identify potential galaxy-galaxy lenses, investigate multi-colour images available from ground observations of this sub-sample and finally subtract the lens light to verify the lens configuration.
Since the first steps in the lens search are based on observation in a single photometric image, this sample offers a perfect comparison to test our model predictions.
We compare both the full sample of 67 candidate lenses presented in \cite{Faure_2008} and the sub-sample with the highest probability (20 lenses).
In order to reproduce the limits imposed by the targeted search in \cite{Faure_2008}, we initialize our model with a restricted range of possible lens redshifts $0.2 < z_l < 1$ and values of velocity dispersion $\sigma > 160$ km/s.
We obtain the cut in velocity dispersion from the magnitude cut via the $M_V - \sigma$ relation from \cite{M_v_vel_disp_relation}.
We use this same $M_V - \sigma$ relation to obtain the distribution of lens magnitudes in the i-band, after accounting for the apparent magnitude accounting for the distance modulus and K-correction.
The left panel in Figure \ref{fig:COSMOS_HST_Faure} shows a good match between the lens redshift distributions of the observed samples and the model, other than a second peak found in the full sample around $z=1$. 
This difference might be due to a bias in the visual selection of candidates, or the variance in the photometric redshift estimation. 
In the central panel, we see that the model Einstein radii distribution matches that of the full sample well, while the smaller sample of secure lenses is skewed towards larger radii. This might be traced back to the fact that larger Einstein radii make lenses easier to identify by visual inspection and hence be classified as highly probable lenses.
Similarly, the apparent magnitude distribution of the lenses in the right panel is skewed towards brighter lenses (i.e., with a larger Einstein radius).
Our model predicts 21 (7) lenses, or 13 (4) detectable through the lens light, assuming the VDF from \cite{Mason_2015} (\citealt{Geng_2021}), imposing the lens search observational constraints within the COSMOS HST area.
It is important to note that the cosmic variance might play a significant role over a small field: using the \textsc{Cosmic Variance Calculator} from \cite{TrentiCosmic_Variance} we find that over this field we should expect uncertainty in the number counts of $\approx 24\%$, accounting for the Poisson error over 67 galaxies and the cosmic variance of within the pencil beam up to $z_l = 1$ (this becomes $\approx 40\%$ if we consider the sample of 20 high probability candidates). This would make both VDF models compatible with the best sample while favouring a model based on the \cite{Mason_2015} VDF if all the candidates in the full sample were genuine lenses.

\subsection{Strong Lensing Legacy Survey (SL2S)}

\begin{figure*}
  \includegraphics[width=\linewidth]{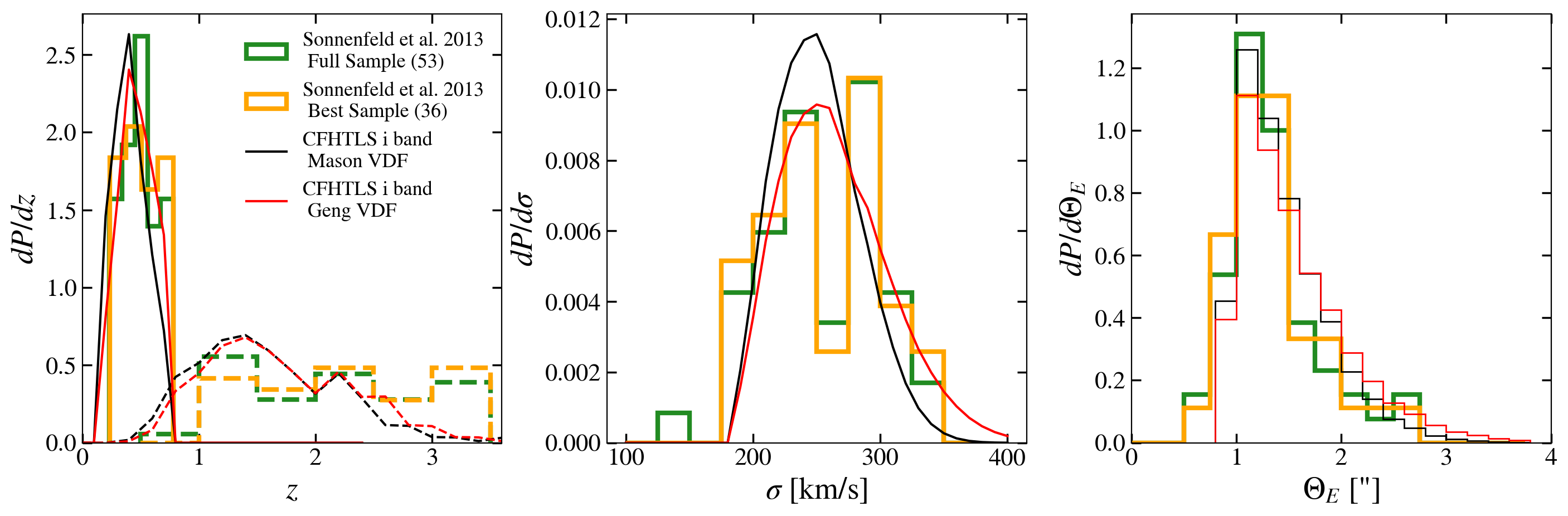}
    \caption{The SL2S lens sample statistics (full sample in thick green line, best sample in thick orange line) compared to our model expectations for the CFHTLS $i$ band assuming the VDF from \citealt{Mason_2015} (in black) or \citealt{Geng_2021} (in red). 
    The left panel shows the redshift distributions of the lens (solid) and source (dashed) populations. The middle panel shows the velocity dispersion distributions. The right panel shows the distributions of Einstein radii.
    The data bin width is based on the Freedman–Diaconis rule.}
    \label{fig:SL2S_Sonnenfeld}
\end{figure*}

The Strong Lensing Legacy Survey (SL2S, \citealt{CFHTLS_SL2S}) was designed to find strong lens systems in the Canada–France–Hawaii Telescope Legacy Survey (CFHTLS). 
The targeted lens search procedure in the SL2S is described in \cite{Gavazzi_SL2S_2012} and uses the \textsc{ringfinder} algorithm\footnote{This technique is also used in \cite{Collett_2015} as one of the two detectability criteria for galaxy scale lenses (the other being the simultaneous satisfaction of multiple images, resolved Einstein ring, pronounced tangential arcs and minimum SNR requirements).} (\citealt{RINGFINDER_ALGORITHM}), which imposes a limit in magnitude ($i<22.5$) and redshift ($0.1 < z_l < 0.8$) of the lens, and then looks for blue rings by subtracting a scaled, point-spread function (PSF) matched version of the $i$-band image from the $g$-band image.
The lens candidates found in the CFHTLS photometry using this algorithm have been followed up with higher-resolution photometric observations using HST and spectroscopy to determine the redshift and velocity dispersion of the galaxies in the sample.
With this setup, \cite{Gavazzi_SL2S_2012} find 2-3 lens candidates per square degree, and \cite{Sonnenfeld_SL2S_2013} reports 36 secure lenses and 17 lens candidates (for which no spectroscopy was available).
We initialize our model with the same restricted range lens redshifts used in the search and velocity dispersion $\sigma > 180$ km/s, obtained from the cut in i-band flux imposed via the K corrected $M_V - \sigma$ relation from \cite{M_v_vel_disp_relation}. 
In Fig. \ref{fig:SL2S_Sonnenfeld} we compare the catalogue of lenses properties in SL2S (\citealt{Sonnenfeld_SL2S_2013a}, \citealt{Sonnenfeld_SL2S_2013}) to the lens statistics predicted by our model for this field, imposing the same constraints on the redshift and velocity dispersion ranges over the CFHTLS Wide field (which observed 171 deg$^2$ down to $i \approx 24.5$).
We compare both the full sample of 53 candidate lenses with available spectroscopic information and the sub-sample of 36 grade A lens candidates.
The left panel shows good agreement between the lens redshift distributions of the observed samples and the models, even though the redshift distributions of the background sources appear flatter in the observed distributions, leading to a greater fraction of high redshift sources. 
The central and right panels show excellent agreement between the velocity dispersion and Einstein radius distributions, respectively. 
The disagreement of the Einstein radii distribution is consistent with a source redshift distribution skewed towards higher values of $z$ since, for a fixed lens population at centred around $z_l \approx 0.5$, the Einstein radius $\Theta_E$ of a source at redshift $z_s\approx3.5$ would be around 25\% larger than $\Theta_E$ of a source at redshift $z_s\approx1.5$.
Accounting for the constraints on the lens redshift range from \cite{Gavazzi_SL2S_2012} and the $\approx25\%$ completeness reached by the \textsc{ringfinder} algorithm after visual inspection for magnifications $\mu > 4$ (\citealt{RINGFINDER_ALGORITHM}), our model predicts $174$ ($51$) lenses within the CFHTLS Wide area, or $92$ ($46$) detectable through the lens light, assuming the VDF from \cite{Mason_2015} (\citealt{Geng_2021}).

\subsection{Machine learning based search in DES}
\begin{figure}
  \includegraphics[width=\linewidth]{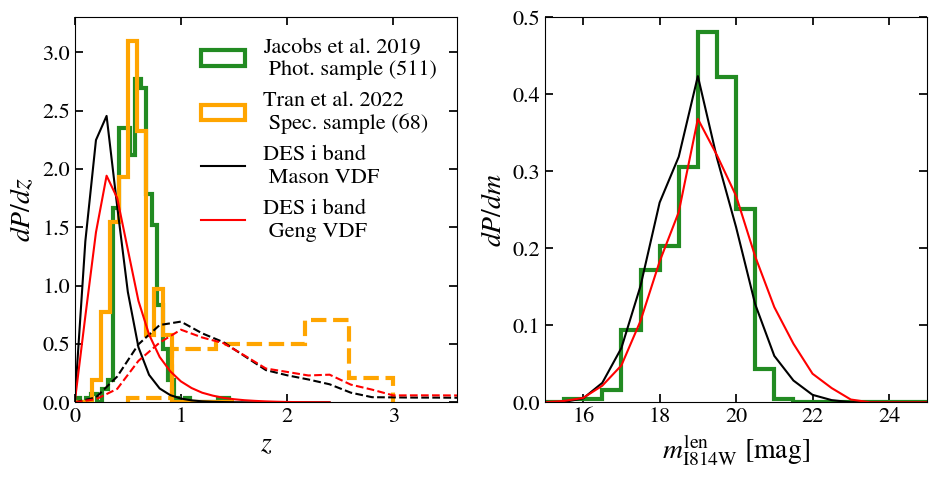}
    \caption{The lens sample statistics from a machine learning based lens search on the DES data (\citealt{Jacobs_catalog}, thick green line) and its spectroscopic follow-up (\citealt{AGEL_Survey}, thick orange line), compared to our model expectations for the DES $i$ band assuming the VDF from \citealt{Mason_2015} (in black) or \citealt{Geng_2021} (in red).
    The left panel shows the redshift distributions of the lens (solid) and source (dashed) populations. The right panel shows the distributions of apparent magnitudes of the lens sample.
    The data bin width is based on the Freedman–Diaconis rule.}
    \label{fig:Jacobs_DES}
\end{figure}

The lens search based on a CNN architecture in \cite{Jacobs_catalog} was designed to classify the potential lenses in the DES first data release (\citealt{DES_3yr_DataRelease}).
The training sample for the neural network was generated from the \textsc{lenspop} code described in \cite{Collett_2015}.
The network is trained on a subsample of the lenses that \textsc{lenspop} forecasts for DES designed to maximize the difference between the lens and non-lens populations, by tightening the thresholds used for detectability: SNR$_\text{min} = 20$, $\mu_\text{arc} = 5$, and an Einstein radius $>2$".
It is important to note that a neural network trained on a certain volume of the parameter space, could classify objects outside the parameter range it was trained on (e.g., see the redshift range of lenses in \citealt{Jacobs_CNN_DES}).
In general, it is therefore impossible to predict how the threshold imposed on the training sample will translate on to the final candidate selection.
A follow-up study on the sensitivity of this CNN architecture performance on the degradation of the input parameters (\citealt{Jacobs_CNN_parameters_for_lensing}) found that the networks are highly sensitive to colour, the simulated PSF used in training, and occlusion of light from a lensed source, but are insensitive to Einstein radius, and performance degrades smoothly with source and lens magnitudes.
In Fig. \ref{fig:Jacobs_DES} we compare the catalogues of lenses from \cite{Jacobs_catalog} and its spectroscopic follow-up, the AGEL survey (\citealt{AGEL_Survey}), to the lens statistics predicted by our model for the DES field in the $i$ band (which observed $\sim 5000$ deg$^2$ down to $i \approx 23$). 
The left panel shows that the observed candidate lens distribution is characterized by a higher average redshift compared to our model expectations, is still present in the AGEL survey (see Fig. 7 in \citealt{AGEL_Survey}).
This might suggest that either there is a bias in the lens selection algorithm, or the underlying VDF has to be skewed to higher redshifts.
The fact that the apparent magnitudes of the lenses are roughly consistent with our model (right panel), might indicate that the shift in redshift distribution is not due to the VDF profile.
The lens search in \cite{Jacobs_catalog} yielded 511 lens candidates, while the AGEL sample reports 68 spectroscopically confirmed lenses out of 77 systems selected, giving a successful lens identification rate of $\approx88\%$.
For this field, our model predicts 1822 (440) lenses, or 828 (269) detectable through the lens light, assuming the VDF from \cite{Mason_2015} (\citealt{Geng_2021}). 
Imposing the stricter conditions applied to the training set of the ML model, we expect 479 (139) lenses, or 180 (75) visible through the lens light, using the \cite{Mason_2015} (\citealt{Geng_2021}) VDF.
On the other hand, assuming that the 88\% purity reported in \cite{AGEL_Survey} is constant over the whole ML selected sample, we should expect around $450$ lenses in the complete AGEL sample, challenging the VDF redshift evolution in the models considered so far.
The choice of the machine learning model and architecture has a strong impact on the proportion of true positives and false positives, which is a measurable quantity in the training and validation process, but could also introduce a bias on the lens and source statistics which is very difficult to quantify.
This work could be useful to generate a set of samples drawn from distributions resulting from a range of input parameter values and use these to probe the response of machine learning algorithms.
\REPLY{
In Appendix A we find a set of VDF parameters that produce distributions compatible with the outcomes of the machine learning based search in DES. We find that a strong evolution of the slope $\alpha$ and characteristic velocity dispersion $\sigma_\star$ are necessary to match these observations. Such evolution is hard to reconcile theoretically to the observed luminosity function evolution.
}

\subsection{SuGOHI}
\begin{figure}
  \includegraphics[width=\linewidth]{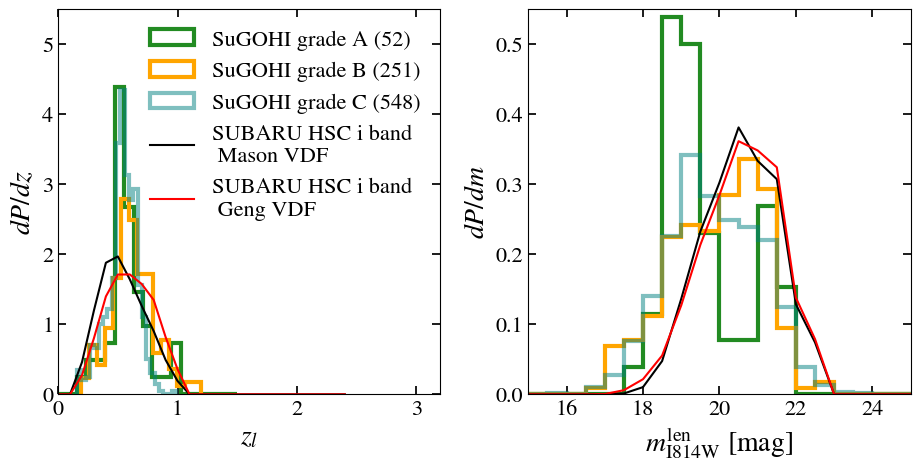}
    \caption{The lens sample statistics from the SuGOHI lens catalogue (\citealt{Sonnenfeld_SuGOHI_I} and following papers), divided by the lens grade (A in green, B in orange and C in cyan), compared to our model expectations for the DES $i$ band assuming the VDF from \citealt{Mason_2015} (in black) or \citealt{Geng_2021} (in red).
    The left panel shows the redshift distributions of the lens (solid) and source (dashed) populations. The right panel shows the distributions of apparent magnitudes of the lens sample.
    The data bin width is based on the Freedman–Diaconis rule.}
    \label{fig:SUGOHI}
\end{figure}

The Survey of Gravitationally-lensed Objects in HSC Imaging (SuGOHI) is an ongoing series of lens searches on the Hyper Suprime-Cam Subaru Strategic Program (HSC SSP, \citealt{SUBARU_HSCSSP}) survey, based on a variety of different searching techniques.
In this section, we make use of the SuGOHI Candidate List\footnote{\url{https://www-utap.phys.s.u-tokyo.ac.jp/~oguri/sugohi/}}, limiting the analysis to galaxy-galaxy lenses grouped by their grade.
The catalogue is constructed combining three lens search methods: feature detection algorithms and model fitting (e.g., \textsc{yattalens} in \citealt{Sonnenfeld_SuGOHI_I}, \citealt{Wong_SuGOHI2}, \citealt{Wong_SUGOHI}), human inspection (\citealt{Jaelani_SuGOHI5} and \citealt{Sonnenfeld_citizien_science}) and Machine Learning (\citealt{Jaelani_SuGOHI10}).
All of these searches are based on different data releases of the HSC SSP survey, but since they mainly differ in the area covered by the data sample we choose to compare them to the probability distribution based on our model of the final data release of HSC SSP (which observed $1400$ deg$^2$ down to $i \approx 26.2$).
Furthermore, each SuGOHI search features unique selection criteria on the sample of galaxies inspected as lens candidates. In general, all of them restrict the lens search to massive ETGs ($\sigma \gtrsim 230$ km/s) with redshift within $0.2 \lesssim z_l \lesssim 1.0$.
After imposing these constraints on our model, we compare it with the properties of the SuGOHI Candidate List as shown in Fig. \ref{fig:SUGOHI}. 
In the left panel, the observed candidate lens distribution appears more peaked around its central value than our model prediction. This might reflect the prior distribution of luminous red galaxies (LRGs) selected from the Baryon Oscillation Spectroscopic Survey (BOSS) sample as the input sample for the lens search in many of the searches composing the galaxy-galaxy lens catalogue (e.g., see Fig. 7 in \citealt{Sonnenfeld_SuGOHI_I}). 
The lens magnitude distributions shown in the right panel of Fig. \ref{fig:SUGOHI} show good agreement between the model and the observations, in particular for the grade B 'probable' lenses. 
Grade A lenses appear to be skewed towards brighter lenses, with larger Einstein radii.
For this field, our model predicts $\approx 10^4$ lenses assuming the VDF from \cite{Mason_2015}, and $\approx 5 \times 10^3$ lenses assuming the \citealt{Geng_2021} VDF. 
Since the sample is constructed from different data releases and slightly different pre-selection criteria, we decided to not compare the number of lenses to our models.

\subsection{Summary and interpretation of comparisons}

In this section, we have compared the results of our model to four lens searches.
In all the comparisons we showed the distributions predicted by our model for two different choices of VDF, after applying the constraints on the parameter space (e.g., redshift, magnitude, or Einstein radius of the lens) adopted in each lens search.
The results of these models are listed in Table \ref{tab:Past_Searches_results}.

Our model shows good agreement between the lens apparent magnitude (obtained from the lens velocity dispersion) and Einstein radii distribution across all surveys. 
The predicted lens and source redshift distributions match well with the results of the COSMOS HST and SL2S surveys, but show some tension with the redshift distributions in SuGOHI and the outcomes of the ML-based search on DES data. 
These differences might be traced back to the difference in the completeness of the search methods over the range of redshift considered, even though further work is needed to constrain the VDF evolution with cosmic time and rule out the possibility of observed redshift distributions being linked to the lens population.
The number of lenses predicted by our model agrees with the observed number counts in \cite{Faure_2008} and \cite{AGEL_Survey}, after imposing the constraints on $z_l$, $\sigma$, $z_s$, and $\Theta_E$ of the related lens search method.
This agreement provides additional validation of our model.
In fact, \cite{Faure_2008} find $20$ probable lenses in COSMOS HST, and our model predicts between $4$ and $21$ identifiable lenses in the same field, depending on the assumed VDF and the ability to remove the lens light.
Similarly, given the purity of the \cite{AGEL_Survey} we can expect around $450$ lenses in DES, which is within our predicted range of $75$ to $479$ lenses that can be found within the constraints of their search method. 
Using the same parameter range of the lens population as imposed in \cite{Sonnenfeld_SL2S_2013a}, we predict a total of $46$ to $174$ lenses in the CFHTLS.
Since the SuGOHI Candidate List is constructed using different data releases (i.e., sky areas) and different search methods, we do not compare its number of lenses to our predictions.

The comparison with past lens searches conducted in this Section shows that our model gives a reliable estimate of the number and distribution of identifiable lenses in surveys.
We therefore turn to predictions for future surveys in the following section.

\begin{table*}
  \caption{Galaxy-galaxy lensing yields an estimate for a selection of past lens searches.}
  \label{tab:Past_Searches_results}
  \centering
  \begin{tabular*}{\textwidth}{@{\extracolsep{\fill}} c c c c c c c c c c c c }
      \hline 
      Telescope  & Survey      & Filter & Seeing (PSF) & Area      & Exp. time & $m_{\text{cut}}$ & $m_{\text{lim}}$ & $N$          &  $N$         & $N$               & Ref.          \\
                 &             &        & ["]          & [deg$^2$] & [sec]     & [mag]            & [mag]            & VDF M        &  VDF G       & [deg$^{-2}$]      &               \\
          (1)    &     (2)     &   (3)  & (4)          & (5)       & (6)       & (7)              & (8)              & (9)          &  (10)        & (11)              & (12)          \\
      \hline 
      \hline 
      HST        & COSMOS      & F814W  & 0.120        & 1.60      &  2000     & 25.0             & 26.5             &   36 (17)    &    11 (7)    &    10 | 4         & \scriptsize$1$ \\
                 & COSMOS (F)  &        &              &           &           &                  &                  &   21 (13)    &    7  (4)    &     8 | 2         & \\
      CHFT       & CHFTLS      & i      & 0.620        & 170       &  5500     & 24.5             & 25.1             &  975 (525)   &   268 (185)  &     3 | 1         & \scriptsize$2$ \\
                 & CHFTLS (S)  &        &              &           &           &                  &                  &  174 (92)    &   51 (46)    &  0.54 | 0.27      & \\
      DECam      & DES         & i      & 0.960        & 5000      &  900      & 23.0             & 24.7             & 1822 (828)   &   440 (269)  &  0.17 | 0.05      & \scriptsize$3$ \\
                 & DES (J)     &        &              &           &           &                  &                  & 479  (180)   &   139 (75)   &  0.04 | 0.02      & \\
      HSC        & HSC-SSP     & i      & 0.600        & 1400      &  8000     & 26.2          & 26.2   & 5.2 (2.8) $\times10^4$ &  1.6 (1.2)  $\times10^4$ &   20 | 9 & \scriptsize$4$ \\
                 & HSC-SSP (Su)&        &              &           &           &               &        & 2.3 (1.7) $\times10^4$ &  0.9 (0.7)  $\times10^4$ &   12 | 5 & \\
      \hline   

  \end{tabular*}
  \begin{flushleft}
  \footnotesize \textit{Notes.} 
  \footnotesize The columns indicate the (1) telescope, (2) survey, and (3) photometric filter considered. (4) The seeing (or PSF for space-based telescopes) in arcseconds. 
  \footnotesize (5) The area of a survey in square degrees. (6) The exposure time in seconds. (7) The magnitude cut and (8) the limit magnitude. 
  \footnotesize (9 - 10) The total number of identifiable lenses, with lens light (not) removed, assuming the VDF described in \citealt{Mason_2015} and \citealt{Geng_2021}, respectively.
  \footnotesize (11) The surface number density, in square arcseconds, of the lenses identifiable through the foreground lens light, 
  \footnotesize assuming the VDF described in \citealt{Mason_2015} and \citealt{Geng_2021}, respectively. (12) The references for the input parameters.
  \footnotesize When an initial is indicated in parenthesis, as in (F), it indicates a set of additional constraints on the lens and/or source distribution imposed by a particular lens search. The 
  \footnotesize (F) refers to the lens search in the COSMOS field by \citealt{Faure_2008},
  \footnotesize (S) to the lens search in the CHFTLS field by \citealt{Sonnenfeld_SL2S_2013a},
  \footnotesize (J) to the lens search in the DES field by \citealt{Jacobs_CNN_DES}, and
  \footnotesize (Su) to the lens search in the HSC-SSP field by \citealt{Sonnenfeld_SuGOHI_I} and following SuGOHI papers.\\
  \footnotesize \textit{Ref.}
  \footnotesize $^1$ \citealt{HST_COSMOS},
  \footnotesize $^2$ \citealt{CFHTLS_SL2S},
  \footnotesize $^3$ \citealt{DES_3yr_DataRelease},
  \footnotesize $^4$ \citealt{SUBARU_HSCSSP}.
  \end{flushleft} 
\end{table*}

\section{Forecast of strong lens yields in surveys}

\begin{figure*}
  \includegraphics[width=\linewidth]{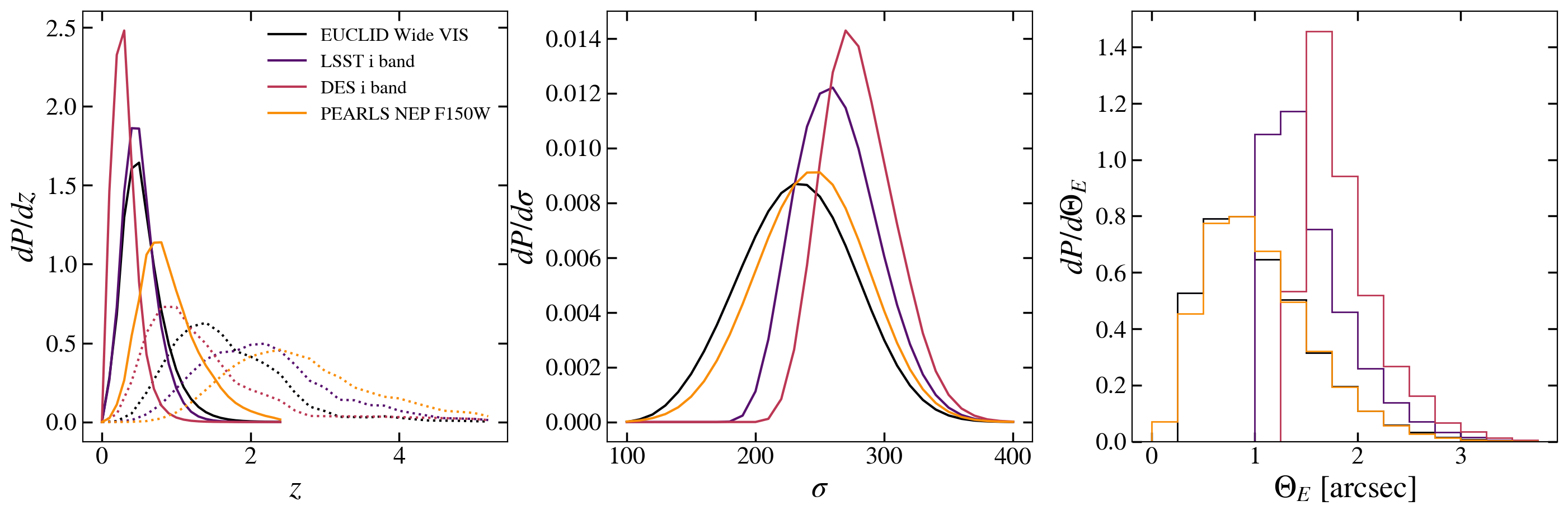}
    \caption{Properties of the forecasted lens samples in a selection of wide and narrow field surveys, adopting the \citealt{Mason_2015} VDF. 
    The left panel shows the redshift distributions of the lens (solid) and source (dotted) populations.
    The middle and right panels show the velocity dispersion and the Einstein radius distributions, respectively.
    Black shows Euclid, purple shows Vera Rubin LSST, red shows DES and orange shows the JWST PEARLS NEP field.}
    \label{fig:comp_surveys}
\end{figure*}

\begin{figure*}
  \includegraphics[width=\linewidth]{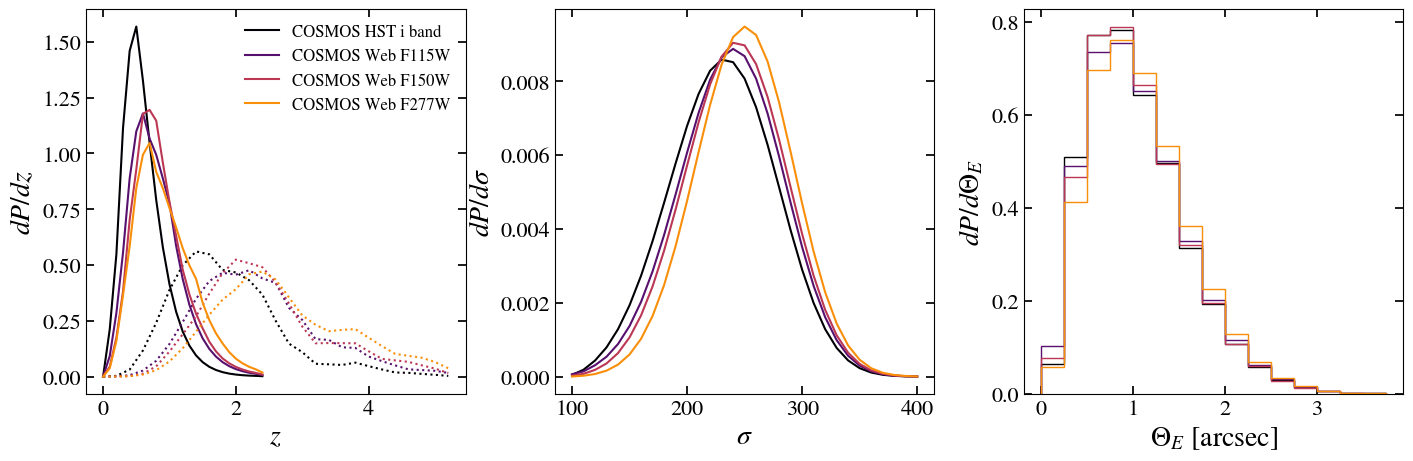}
    \caption{
    \REPLY{Properties of forecasted lens samples in a selection of three JWSTs and one HST photometric bands.
          Black shows $i$ band (HST), purple shows F115W, red shows F150W and orange shows F277W (JWST).
          Same panel composition as in Fig. \ref{fig:comp_surveys}}}
    \label{fig:comp_bands}
\end{figure*}

In this section we investigate the galaxy scale lens and source populations in a range of ground-based and space-based surveys, encompassing a range of areas, depth and photometric bands, as listed in Table \ref{tab:Survey_results}. 
We forecast the expected number of lenses assuming that the lens light can (not) be completely removed, which gives us an upper (lower) limit on the lens yield of a given survey. We also compute the lens statistics for each survey using two different choices of VDF (\citealt{Mason_2015} and \citealt{Geng_2021}). 
We then compare the distributions of three surveys (EUCLID Wide, LLST, DES) and the JWST PEARLS NEP field in Fig. \ref{fig:comp_surveys}.
\REPLY{From this comparison, one can see that the main factors contributing to the forecasted distributions and the total number of detectable lenses are:}
\begin{itemize}
\item \REPLY{the photometric band in which we conduct the observations (although part of the differences between bands is due to the change in survey depth).}
For example sources with redshift $z_s < 5.3$ can be detected in the $i$-band, while sources up to $z_s \lesssim 12$ can be seen in the F150W band.
\REPLY{In Fig. \ref{fig:comp_bands} we show the distributions for four bands over the COSMOS field: HST's $i$ band, and JWST's F115W, F150W, and F277W.}
\item \REPLY{The flux limit combined with the luminosity density of the source population, that gives an upper bound on the maximum redshift detectable.}
This is why in Fig. \ref{fig:comp_surveys} the redshift distribution of the JWST PEARLS NEP field, which is extremely deep, is skewed to higher values of both lens and source redshift compared to the distributions of the other surveys shown in the same panel.
\item \REPLY{The PSF (or seeing), which sets the smallest Einstein ring that can be resolved.}
\REPLY{This explains why} the final co-added sample of lenses from LSST is expected to have a lower central value for the lens redshift distribution and a comparatively high average source redshift.
In fact, for an isothermal lens, the Einstein radius is proportional to the ratio of angular diameter distances $D_{ls}/D_s$, and hence surveys with large seeing/PSF (e.g., ground-based imaging) can not probe high redshift and low mass deflectors, as they have smaller Einstein radii (for fixed source redshift).
For example, in Fig. \ref{fig:comp_surveys} we can see how the distributions of Einstein radii in DES are characterized by a larger lower bound compared to LSST observed in the same band.
This drives DES to have a lens redshift distribution peaked around a smaller value of $z_l$ and a lens velocity dispersion distribution centred around a higher value of $\sigma$ compared to the lens population observable by LSST.
\end{itemize}
This shows that the properties of ETGs that are most likely to be a lens depend on the characteristics of the telescope and survey.
Hence lens-based searches (i.e., searches that look for lenses around a pre-selected sample of ETGs), should choose the selection criteria for the potential lens population after simulating the expected distributions in parameter space to minimize the loss of identifiable lenses.

We find that Euclid, Roman and Vera Rubin telescope wide surveys will yield $\mathcal{O}(10^5)$ lenses, with the exact number varying by up to a factor of $4$ depending on the ability to remove the lens light and the lens mass distribution (or VDF).
The predicted surface density of lenses identifiable through the lens light (indicated in Table \ref{tab:Survey_results}) shows a wide range of expected values. 
For example, the Euclid Wide survey and a single visit of LSST should expect a few lenses per square degree, while a deep JWST field should expect around $50$ to $100$ lenses per square degree.
More details for some combinations of surveys and observing photometric bands can be found in Table \ref{tab:Survey_results}.

The comparisons conducted in the previous section might suggest that the actual number of identifiable lenses can be significantly smaller, depending on the search method and specific constraints applied to the observed data.

\REPLY{
In Appendix B we briefly compare our results with the ones from \citealt{Collett_2015} and the Euclid Collaboration.
}

\begin{table*}
  \caption{Galaxy-galaxy lensing yields estimate for a selection of ongoing and future surveys observing in the visible/NIR.}
  \label{tab:Survey_results}
  \centering
  \begin{tabular*}{\textwidth}{@{\extracolsep{\fill}} c c c c c c c c c c c c }
      \hline 
      Telescope  & Survey      &Filter & Seeing (PSF)   & Area      & Exp. time & $m_{\text{cut}}$  &$m_{\text{lim}}$& $N$                    &  $N$                   & $N$          &  Ref. \\
                 &             &       & ["]            & [deg$^2$] & [sec]     & [mag]             & [mag]          & VDF M                  &  VDF G                 & [deg$^{-2}$] &       \\
      \hline 
      \hline 
                 &             & F115W & 0.040          & 0.54      &  516      & 26.13             & 27.13          & 46 (15)                & 16 (8)                 &    28 | 15   &       \\
                 & COSMOS Web  & F150W & 0.050          & 0.54      &  516      & 26.35             & 27.35          & 74 (14)                & 25 (8)                 &    26 | 15   & \scriptsize$1$\\
                 &             & F277W & 0.092          & 0.54      &  516      & 27.00             & 27.99          & 121 (15)               & 42 (10)                &    28 | 19   &       \\
      JWST       &             &       &                &           &           &                   &                &                        &                        &              &       \\
                 &             & F115W & 0.040          & 0.015     &  11680    & 27.76             & 28.76          & 4 (1)                  & 1 (1)                  &    57 | 31   &       \\
                 & PEARLS NEP  & F150W & 0.050          & 0.015     &  11680    & 27.91             & 28.91          & 8 (1)                  & 3 (1)                  &    73 | 46   & \scriptsize$2$\\
                 &             & F277W & 0.092          & 0.015     &  11680    & 27.81             & 28.81          & 7 (1)                  & 3 (1)                  &    43 | 31   &       \\
      \hline
      Euclid     & Wide        & VIS   & 0.170          & 15000     &  2280     & 24.50             & 24.50          & 1.5 (0.9) $\times10^5$ & 4.5 (3.2) $\times10^4$ &     6 | 2    & \scriptsize$3$\\
                 &             & Y     & 0.220          & 15000     &  448      & 24.00             & 24.00          & 0.5 (0.2) $\times10^5$ & 1.3 (0.7) $\times10^4$ &     1 | 1    &       \\
                 &             & J     & 0.300          & 15000     &  448      & 24.00             & 24.00          & 0.8 (0.4) $\times10^5$ & 2.4 (1.4) $\times10^4$ &     2 | 1    &       \\
                 &             & H     & 0.360          & 15000     &  448      & 24.00             & 24.00          & 1.2 (0.4) $\times10^5$ & 3.3 (1.5) $\times10^4$ &     2 | 1    &       \\
      \hline 
      Roman      &High Latitude& J129  & 0.300          & 1700      &  146      & 26.7              & 26.7           & 4.5 (1.7) $\times10^4$ & 1.4 (0.8) $\times10^4$ &    10 | 5   & \scriptsize$4$\\
                 &Wide-Area    &       &                &           &           &                   &                &                        &                        &              &       \\
      \hline  
      Vera Rubin& single visit & i     & 0.710          & 20000     &  30       & 23.41             & 23.41          & 2.2 (1.3) $\times10^4$ & 5.7 (4.2) $\times10^3$ &     1 | 0.24 & \scriptsize$5$ \\
      (LSST)    &final co-added& i     & 0.710          & 20000     &  6000     & 26.40             & 26.40          & 7.4 (3.8) $\times10^5$ & 2.4 (1.7) $\times10^5$ &    19 | 8    & \\
      \hline   
  \end{tabular*}
  \begin{flushleft}
  \footnotesize \textit{Notes.} This Table follows the same structure as Table \ref{tab:Past_Searches_results}.\\
  \footnotesize \textit{Ref.} $^1$ \citealt{COSMOS_JWST},
  \footnotesize$^2$ \citealt{Windhorst_2022_PEARLS},
  \footnotesize$^3$ \citealt{Euclid_Wide_Survey},
  \footnotesize$^4$ \citealt{Roman_WFIRST},
  \footnotesize$^5$ \citealt{LSST_Ivezic_2019}.
  \end{flushleft}
\end{table*}

\section{Rare lenses}

The large samples of strong lenses that will be discovered with upcoming wide area surveys will contain rare configurations such as double plane systems and quad lenses.

\subsection{Dual lenses}

A single galaxy might act as a lens for more than one background source, in general at different redshifts (excluding two merging galaxies as the source).
This multiple lens plane configuration might be used as a cosmography tool to probe global cosmological parameters such as $\Omega_m$ or $\Omega_\Lambda$, via the ratio of cosmological distances entering the lensing potential (e.g. \citealt{Jackpot_Gavazzi}).

An order of magnitude estimate of the probability of observing multiple sources behind a single lens can be obtained by applying Poisson statistics to the \textit{a-priori} probability $\tau(z_s)$ of lensing a source at redshift $z_s$.
Defining $P(N)$ as the probability of detecting a lens with $N$ source planes, we should expect around $P(2)/P(1)=\tau/2!$ double lenses per single lens, or one in $\sim 10^3$.

For a more thorough calculation, assuming the lensing cross sections of the two populations of sources are independent (\citealt{Jackpot_Gavazzi}), we can write the number of double lenses as

\begin{equation}\label{eq:Number_dbl_lens_galaxies}\begin{aligned}
  &N_{s1, s2} =  
  \int \dd{z_{s2}} 
  \int \dd{z_{s1}} 
  \int \dd{M_{s2}}
  \int \dd{M_{s1}} 
  \int \dd{z_l} 
  \int \dd{\sigma} 
  \Phi(\sigma, z_l)\\
  &\times \frac{\dd N_{SL}^{obs}(M_{s1}, z_{s1} | \sigma, z_l)}{\dd z_{s1} \dd M_{s1}}
  \frac{\dd N_{SL}^{obs}(M_{s2}, z_{s2} | \sigma, z_l)}{\dd z_{s2} \dd M_{s2}}
  \dv{V(A_s, z_l)}{z_l} 
  \:.
\end{aligned}\end{equation}

\begin{figure}
  \includegraphics[width=\linewidth]{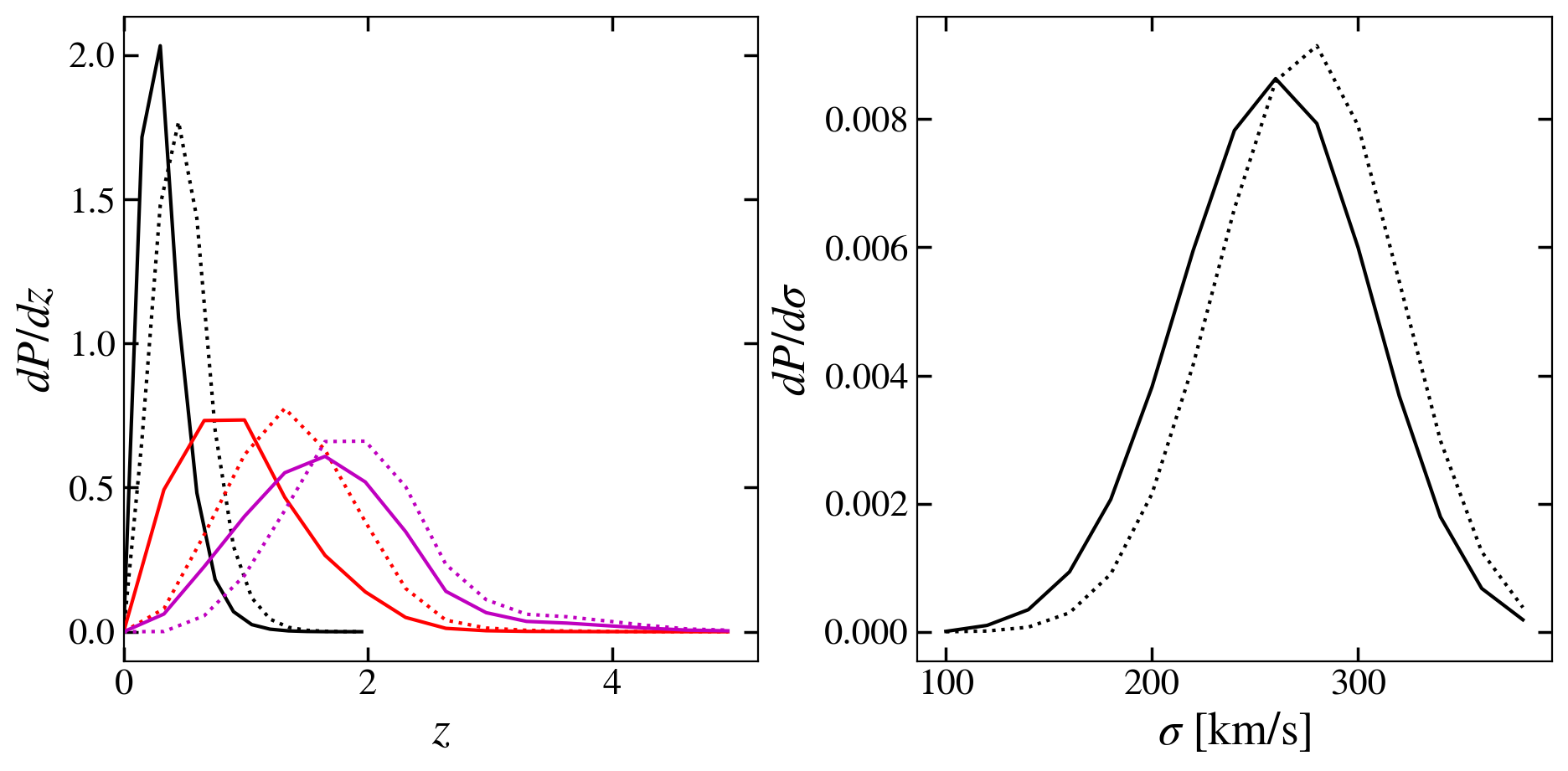}
    \caption{Lens and sources redshift (left) and velocity dispersion (right) distributions for the population of detectable dual plane lenses. In the left panel, the lens redshift is shown in black, the first source redshift is red, and the second source redshift is magenta. The solid (dotted) line represents the lenses detectable when the source light is (not) removed.}
    \label{fig:Euclid_dbl_lenses}
\end{figure}

Applying Eq. \ref{eq:Number_dbl_lens_galaxies} to Euclid Wide VIS, we predict 154 (28) double lenses, assuming lens light can (not) be removed (see Fig. \ref{fig:Euclid_dbl_lenses}).
This implies a fraction of double lenses of $1\times 10^{-3}$ ($0.3\times 10^{-3}$). 
Increasing the depth of the survey yields a higher fraction of double lenses, up to a fraction of $\sim 3\times 10^{-3}$.
A ratio of a double lens every thousand lenses is compatible with observations, given that so far have found only a handful of double lenses (\citealt{Jackpot_Gavazzi}, \citealt{Tu_double_lens}, \citealt{Tanaka_double_lens}, \citealt{Schuldt_double_lens}) out of the $\sim 10^3$ discovered lenses.

\subsection{Quads}

Adopting the formalism developed in Section 2.3, we find that $\sim4$\% of the lenses should be in a quad configuration, which is lower than previous estimates on galaxy-quasar lenses (e.g., $\sim10\%$ in \citealt{Oguri_Marshall}, and $9\%$ \citealt{Sonnenfeld_SL_selection}).
However, our value is consistent with existing observations of galaxy-galaxy (e.g., 5\% in \citealt{More_citizien_science}).
The differences between these estimates are due to the different choices in the lens mass distribution and its evolution with redshift, in the distribution of axis ratio in lenses, and in the luminosity density of sources compared to the survey limit.
The estimate of \cite{Oguri_Marshall} is based on a SIE lens with an external shear component, and normally distributed ellipticities centred on an axis ratio of $f=0.7$.
\citealt{Sonnenfeld_SL_selection} finds a quad fraction of 9\%, using a composite lens model (projected Sérsic profile for the stellar component and a projected NFW profile for the dark matter component) and an axis ratio distribution set to a beta distribution with $\alpha = 6.28$ and $\beta = 2.05$.
Quads are biased towards more elliptical lenses (\citealt{Sonnenfeld_SL_selection}), because the size in the source plane of the inner caustic increases with decreasing axis ratio $f$, and larger-sized sources, as they are more likely to overlap with the inner caustic. 
For this reason, the exact value of this fraction comes directly from our assumption about the ellipticity distribution of the lens population. On the other hand, this implies that quads can be used as probes of ellipticity.
The mass distribution of the lens population becomes important when considering extended sources.
We discuss the overall effect of proper modelling of extended sources lensing in the following Section.

\section{Discussion}

This paper has presented an analytic model for lens statistics which adopts a set of prescriptions motivated by observational and theoretical constraints on the deflector and source populations.
In this section we discuss the approximations we used to construct our model, focusing on the regimes where these simplifying assumptions are sufficient to describe the lens statistics and the cases where these assumptions should be refined.

\subsection{Improving the lens model}

We have assumed that the lens population is composed of early-type galaxies with an isothermal total mass profile.
While this assumption is well suited to describe the overall properties of a sample of galaxy scale lenses, different choices of lens mass profiles could affect both the lensing cross-section and the magnification probability distribution.\\
\indent One potential improvement is to sample the deflector population from a simulation rather than model it analytically.
While this gives a much higher level of detail on the projected mass and light distribution of a lens compared to an analytical approximation, it is often limited to a maximum stellar mass given by the simulation volume (e.g., $<10^{11.5}M_\odot$ in \citealt{Holloway_2023}).
This might lead to biases in the lens statistics, especially when compared to lens searches targeted around bright ETGs (e.g. Fig. 7 in \citealt{Sonnenfeld_SuGOHI_I}). 
For example, using the $M_*$-$\sigma$ relation from \cite{smass_veldisp_relation} to map the upper bound in stellar mass to a cut in the VDF, we find that lensing estimators based on realizations of galaxy catalogues might underestimate the amount of low redshift ($0<z\lesssim 0.5$) gal-gal lenses a factor of $\sim 2$, especially in deep surveys.\\
\indent Furthermore, analytical estimates for the lensing cross section as a function of source size and flux, as well as the dark matter fraction and slope for a lens with a generalized Navarro-Frenk-White mass component (\citealt{Wyithe_gNFW}), have been studied in \cite{Sonnenfeld_SL_selection} and could be integrated into our current model.\\
\indent Another improvement would be to include external shear, which can boost the magnification of background sources.
A full characterization of the line-of-sight environment is beyond the scope of this paper, but \cite{Collett_2015} showed that a simple estimate using normally distributed shear components with values compatible to existing models of gal-gal lenses gives a $\sim$5\% boost in the number of lens systems.
The total number of lenses predicted by the model is not very sensitive to small changes in the ellipticity distribution of the deflector population, in agreement with \cite{Collett_2015}, while the number of quad configuration lenses is strongly correlated with the flattening of the lenses.

\subsection{Improving the source model}

Our model does not explicitly account for the differential magnification of extended sources.
Since extended sources are subject to an overall lower magnification, this motivates movement away from analytical models, in favour of simulation-based analyses (\citealt{Collett_2015}).
However, we argue that the inclusion of extended sources has a small effect on the summary statistics of lensing models, although it could be important when considering a sub-sample of highly-magnified sources.
Figure 2 of \cite{Ferrami_Lensing_bright_end} shows that the magnification bias for extended sources is the same as for point-like sources in the power-law regime of the luminosity function.
Hence, our model will be impacted only for sources in the very bright end of the luminosity function, which contribute a negligible fraction of the total number of detectable galaxies (Eqs. \ref{eq:Lens_Fraction} - \ref{eq:Nth_Fraction}). 
Extended sources should also increase our expected fraction of quad configurations, as they would be more likely to overlap with the inner caustic. Moreover, studying the ratio between the area of the inner caustic and the sizes of the source population would be important for study of the statistics of complete Einstein rings.

We can get an upper bound on the effect of extended sources by looking at the integral of the SIE magnification distribution up to a maximum magnification (that depends on the source size).
Approximating sources as circles with uniform luminosity, a galaxy with an angular radius of half the Einstein radius can attain a maximum magnification of $ \mu_\text{max} \approx 4$ (see Fig. 1 in \citealt{Ferrami_Lensing_bright_end}).
We calculate the fraction of lenses with brightest image having $\mu < \mu_\text{max}$ as $\int_2^4 \dd \mu \dv{P}{\mu} = \int_2^4 \dd \mu \frac{2}{(\mu-1)^3} = 8/9$.
This indicates that a population of extremely large sources would at most have a difference in the expected number of lenses of $\approx 12 \%$ (since most of the source population lies in the faint end of the luminosity function)\footnote{Furthermore, the size-maximum magnitude relation evolves quite rapidly. A source with a size of 10\% the Einstein Radius of the lens has a maximum magnification of $\mu_\text{max}\approx10$, which leads to a difference of $80/81$, or less than 2\% difference with a point source population.}. 
In the future, we plan to extend this model to handle extended sources.

\REPLY{The model does not account yet for the blurring between the source and lens light as a function of PSF size.
This would lower the number of lenses seen through the lens light in large PSF/seeing scenarios (e.g. lower the dashed curve in the top panel of Figure 5).}

\subsection{Improving flux limits}

In future work, we plan to explore alternative thresholds for lens detection other than minimum magnification.
While imposing a minimum magnification is a useful approach from a modelling-based perspective, it is non-trivial to determine the distribution of magnifications of a lens sample from observations.  
This is in part due to the mass-sheet degeneracy (\citealt{Mass_sheet_degeneracy}).

\section{Conclusions}

In this paper, we have presented a flexible model for computing the expected population of detectable lenses in photometric surveys.
We tested our model on four samples of lenses from past searches, COSMOS HST, SL2S, SuGOHI, and a lens search in the DES data.

The number of lenses, lens redshift, magnitude, and Einstein radii distributions predicted by our model provide a good match against the lens samples from COSMOS HST and SL2S where completeness and selection can be modelled, and for Einstein radii distributions in all samples.
On the other hand, such differences might also indicate that the properties of the lens population are not well represented by the Velocity Dispersion Functions considered in this paper, and further work will be needed to properly address these tensions.
We then computed the expected number of lenses in several ongoing and future surveys, including Euclid Wide, Vera Rubin Observatory LSST, and the Roman Space Telescope High Latitude Wide Area, along with the forecast lens searches in deeper and smaller areas using JWST photometry.
We found that the assumed velocity dispersion function and the cut in magnitude have the greatest impact on the detectable lens and source populations. 
Our model predicts a yield of $\mathcal{O}(10^5)$ lenses from future wide-area surveys, in agreement with previous estimates.
We estimated the number of double plane lenses in Euclid (around 150), or around 0.1\%, consistent with the available observations. 
This expected increase of two orders of magnitude of the available sample of double lenses will greatly increase their statistical power for cosmography. 
Our model predicts a fraction of quad lenses around 4\%, also consistent with available samples of galaxy-galaxy lenses.
\REPLY{Beyond predicting the number, mass and redshift distributions of lenses in surveys, this model could be used to constrain the VDF evolution with redshift through inference over a sample of lenses with measured redshifts and angular scale.
Furthermore, by comparing lens catalogues extracted from a given field and selected with different methods to the distributions expected from this model it would be possible to gain some insight into the selection biases introduced by a given lens search method.
In particular, it would be possible to test the observational parameters that contribute the most in mapping the model output distributions to the ones found by the different methods.
Finally, this model suggests fiducial values of redshift and mass for the deflector population that would maximize the lens yield in a lens search targeted around bright ellipticals.
}

The code used in this work is made publicly available\footnote{The code is available at \url{https://github.com/Ferr013/GALESS}}.

\section*{Acknowledgments}
\REPLY{We thank the anonymous referee for useful comments that improved the presentation.}
We wish to thank K. Wong, J. Cuby, M. Oguri, Y. Harikane, K. Glazebrook, and C. Jacobs for useful discussions that improved the scope of this work.
This research was supported by the Australian Research Council Centre of Excellence for All Sky Astrophysics in 3 Dimensions (ASTRO 3D), through project number CE170100013.
 
\section*{Data Availability}
The code and the modelled data discussed in this paper are publicly available at \url{https://github.com/Ferr013/GALESS}.

\bibliographystyle{mnras}
\bibliography{Lens_Survey_Stat}

\begin{thebibliography}{}
\makeatletter
\relax
\def\mn@urlcharsother{\let\do\@makeother \do\$\do\&\do\#\do\^\do\_\do\%\do\~}
\def\mn@doi{\begingroup\mn@urlcharsother \@ifnextchar [ {\mn@doi@}
  {\mn@doi@[]}}
\def\mn@doi@[#1]#2{\def\@tempa{#1}\ifx\@tempa\@empty \href
  {http://dx.doi.org/#2} {doi:#2}\else \href {http://dx.doi.org/#2} {#1}\fi
  \endgroup}
\def\mn@eprint#1#2{\mn@eprint@#1:#2::\@nil}
\def\mn@eprint@arXiv#1{\href {http://arxiv.org/abs/#1} {{\tt arXiv:#1}}}
\def\mn@eprint@dblp#1{\href {http://dblp.uni-trier.de/rec/bibtex/#1.xml}
  {dblp:#1}}
\def\mn@eprint@#1:#2:#3:#4\@nil{\def\@tempa {#1}\def\@tempb {#2}\def\@tempc
  {#3}\ifx \@tempc \@empty \let \@tempc \@tempb \let \@tempb \@tempa \fi \ifx
  \@tempb \@empty \def\@tempb {arXiv}\fi \@ifundefined
  {mn@eprint@\@tempb}{\@tempb:\@tempc}{\expandafter \expandafter \csname
  mn@eprint@\@tempb\endcsname \expandafter{\@tempc}}}

\bibitem[\protect\citeauthoryear{{Abbott} et~al.,}{{Abbott}
  et~al.}{2018}]{DES_3yr_DataRelease}
{Abbott} T.~M.~C.,  et~al., 2018, \mn@doi [\apjs] {10.3847/1538-4365/aae9f0},
  \href {https://ui.adsabs.harvard.edu/abs/2018ApJS..239...18A} {239, 18}

\bibitem[\protect\citeauthoryear{{Aihara} et~al.,}{{Aihara}
  et~al.}{2018}]{SUBARU_HSCSSP}
{Aihara} H.,  et~al., 2018, \mn@doi [\pasj] {10.1093/pasj/psx066}, \href
  {https://ui.adsabs.harvard.edu/abs/2018PASJ...70S...4A} {70, S4}

\bibitem[\protect\citeauthoryear{{Alard}}{{Alard}}{2006}]{ARCFINDER_Alard}
{Alard} C.,  2006, \mn@doi [arXiv e-prints] {10.48550/arXiv.astro-ph/0606757},
  \href {https://ui.adsabs.harvard.edu/abs/2006astro.ph..6757A} {pp
  astro--ph/0606757}

\bibitem[\protect\citeauthoryear{{Angora} et~al.,}{{Angora}
  et~al.}{2023}]{Rosati_CNN_search}
{Angora} G.,  et~al., 2023, \mn@doi [\aap] {10.1051/0004-6361/202346283}, \href
  {https://ui.adsabs.harvard.edu/abs/2023A&A...676A..40A} {676, A40}

\bibitem[\protect\citeauthoryear{{Barone-Nugent}, {Wyithe}, {Trenti}, {Treu},
  {Oesch}, {Bouwens}, {Illingworth}  \& {Schmidt}}{{Barone-Nugent}
  et~al.}{2015}]{Barone_Nugent_2015}
{Barone-Nugent} R.~L.,  {Wyithe} J.~S.~B.,  {Trenti} M.,  {Treu} T.,  {Oesch}
  P.,  {Bouwens} R.,  {Illingworth} G.~D.,   {Schmidt} K.~B.,  2015, \mn@doi
  [\mnras] {10.1093/mnras/stv633}, \href
  {https://ui.adsabs.harvard.edu/abs/2015MNRAS.450.1224B} {450, 1224}

\bibitem[\protect\citeauthoryear{{Birrer} et~al.,}{{Birrer}
  et~al.}{2020}]{Birrer_TDCOSMO}
{Birrer} S.,  et~al., 2020, \mn@doi [\aap] {10.1051/0004-6361/202038861}, \href
  {https://ui.adsabs.harvard.edu/abs/2020A&A...643A.165B} {643, A165}

\bibitem[\protect\citeauthoryear{{Blandford} \& {Narayan}}{{Blandford} \&
  {Narayan}}{1992}]{Blandford_Narayan_1992}
{Blandford} R.~D.,  {Narayan} R.,  1992, \mn@doi [\araa]
  {10.1146/annurev.astro.30.1.311}, \href
  {https://ui.adsabs.harvard.edu/abs/1992ARA&A..30..311B} {30, 311}

\bibitem[\protect\citeauthoryear{{Bouwens} et~al.,}{{Bouwens}
  et~al.}{2014}]{Bouwens_UV_cont_slopes}
{Bouwens} R.~J.,  et~al., 2014, \mn@doi [\apj] {10.1088/0004-637X/793/2/115},
  \href {https://ui.adsabs.harvard.edu/abs/2014ApJ...793..115B} {793, 115}

\bibitem[\protect\citeauthoryear{{Bouwens}, {Illingworth}, {Ellis}, {Oesch}  \&
  {Stefanon}}{{Bouwens} et~al.}{2022}]{Bouwens21_data}
{Bouwens} R.~J.,  {Illingworth} G.,  {Ellis} R.~S.,  {Oesch} P.,   {Stefanon}
  M.,  2022, \mn@doi [\apj] {10.3847/1538-4357/ac86d1}, \href
  {https://ui.adsabs.harvard.edu/abs/2022ApJ...940...55B} {940, 55}

\bibitem[\protect\citeauthoryear{{Cabanac} et~al.,}{{Cabanac}
  et~al.}{2007}]{CFHTLS_SL2S}
{Cabanac} R.~A.,  et~al., 2007, \mn@doi [\aap] {10.1051/0004-6361:20065810},
  \href {https://ui.adsabs.harvard.edu/abs/2007A&A...461..813C} {461, 813}

\bibitem[\protect\citeauthoryear{{Cannarozzo}, {Sonnenfeld}  \&
  {Nipoti}}{{Cannarozzo} et~al.}{2020}]{smass_veldisp_relation}
{Cannarozzo} C.,  {Sonnenfeld} A.,   {Nipoti} C.,  2020, \mn@doi [\mnras]
  {10.1093/mnras/staa2147}, \href
  {https://ui.adsabs.harvard.edu/abs/2020MNRAS.498.1101C} {498, 1101}

\bibitem[\protect\citeauthoryear{{Casey} et~al.,}{{Casey}
  et~al.}{2023}]{COSMOS_JWST}
{Casey} C.~M.,  et~al., 2023, \mn@doi [\apj] {10.3847/1538-4357/acc2bc}, \href
  {https://ui.adsabs.harvard.edu/abs/2023ApJ...954...31C} {954, 31}

\bibitem[\protect\citeauthoryear{{Chan}, {Suyu}, {Chiueh}, {More}, {Marshall},
  {Coupon}, {Oguri}  \& {Price}}{{Chan} et~al.}{2015}]{Chan_model_detection}
{Chan} J. H.~H.,  {Suyu} S.~H.,  {Chiueh} T.,  {More} A.,  {Marshall} P.~J.,
  {Coupon} J.,  {Oguri} M.,   {Price} P.,  2015, \mn@doi [\apj]
  {10.1088/0004-637X/807/2/138}, \href
  {https://ui.adsabs.harvard.edu/abs/2015ApJ...807..138C} {807, 138}

\bibitem[\protect\citeauthoryear{{Choi}, {Park}  \& {Vogeley}}{{Choi}
  et~al.}{2007}]{Choi_2007_SDSS}
{Choi} Y.-Y.,  {Park} C.,   {Vogeley} M.~S.,  2007, \mn@doi [\apj]
  {10.1086/511060}, \href
  {https://ui.adsabs.harvard.edu/abs/2007ApJ...658..884C} {658, 884}

\bibitem[\protect\citeauthoryear{{Coe}, {Ben{\'\i}tez}, {S{\'a}nchez}, {Jee},
  {Bouwens}  \& {Ford}}{{Coe} et~al.}{2006}]{Hubble_UDF}
{Coe} D.,  {Ben{\'\i}tez} N.,  {S{\'a}nchez} S.~F.,  {Jee} M.,  {Bouwens} R.,
  {Ford} H.,  2006, \mn@doi [\aj] {10.1086/505530}, \href
  {https://ui.adsabs.harvard.edu/abs/2006AJ....132..926C} {132, 926}

\bibitem[\protect\citeauthoryear{{Collett}}{{Collett}}{2015}]{Collett_2015}
{Collett} T.~E.,  2015, \mn@doi [\apj] {10.1088/0004-637X/811/1/20}, \href
  {https://ui.adsabs.harvard.edu/abs/2015ApJ...811...20C} {811, 20}

\bibitem[\protect\citeauthoryear{{Collett} \& {Auger}}{{Collett} \&
  {Auger}}{2014}]{Collett_Auger_DoubleSrc}
{Collett} T.~E.,  {Auger} M.~W.,  2014, \mn@doi [\mnras]
  {10.1093/mnras/stu1190}, \href
  {https://ui.adsabs.harvard.edu/abs/2014MNRAS.443..969C} {443, 969}

\bibitem[\protect\citeauthoryear{{Collett} et~al.,}{{Collett}
  et~al.}{2018}]{Collett_GR_test}
{Collett} T.~E.,  et~al., 2018, \mn@doi [Science] {10.1126/science.aao2469},
  \href {https://ui.adsabs.harvard.edu/abs/2018Sci...360.1342C} {360, 1342}

\bibitem[\protect\citeauthoryear{{Collett} et~al.,}{{Collett}
  et~al.}{2023}]{4SLSLS_COLLETT}
{Collett} T.~E.,  et~al., 2023, \mn@doi [The Messenger]
  {10.18727/0722-6691/5313}, \href
  {https://ui.adsabs.harvard.edu/abs/2023Msngr.190...49C} {190, 49}

\bibitem[\protect\citeauthoryear{{Cuillandre} et~al.,}{{Cuillandre}
  et~al.}{2024}]{Euclid_Programme_2024}
{Cuillandre} J.~C.,  et~al., 2024, \mn@doi [arXiv e-prints]
  {10.48550/arXiv.2405.13496}, \href
  {https://ui.adsabs.harvard.edu/abs/2024arXiv240513496C} {p. arXiv:2405.13496}

\bibitem[\protect\citeauthoryear{{Despali}, {Vegetti}, {White}, {Giocoli}  \&
  {van den Bosch}}{{Despali} et~al.}{2018}]{Despali_substructure}
{Despali} G.,  {Vegetti} S.,  {White} S. D.~M.,  {Giocoli} C.,   {van den
  Bosch} F.~C.,  2018, \mn@doi [\mnras] {10.1093/mnras/sty159}, \href
  {https://ui.adsabs.harvard.edu/abs/2018MNRAS.475.5424D} {475, 5424}

\bibitem[\protect\citeauthoryear{{Diehl} et~al.,}{{Diehl}
  et~al.}{2017}]{Diehl_DES_search}
{Diehl} H.~T.,  et~al., 2017, \mn@doi [\apjs] {10.3847/1538-4365/aa8667}, \href
  {https://ui.adsabs.harvard.edu/abs/2017ApJS..232...15D} {232, 15}

\bibitem[\protect\citeauthoryear{{Djorgovski} \& {Davis}}{{Djorgovski} \&
  {Davis}}{1987}]{Djorgovski_Davis_1987}
{Djorgovski} S.,  {Davis} M.,  1987, \mn@doi [\apj] {10.1086/164948}, \href
  {https://ui.adsabs.harvard.edu/abs/1987ApJ...313...59D} {313, 59}

\bibitem[\protect\citeauthoryear{{Dressler}, {Lynden-Bell}, {Burstein},
  {Davies}, {Faber}, {Terlevich}  \& {Wegner}}{{Dressler}
  et~al.}{1987}]{Dressler_FP_1987}
{Dressler} A.,  {Lynden-Bell} D.,  {Burstein} D.,  {Davies} R.~L.,  {Faber}
  S.~M.,  {Terlevich} R.,   {Wegner} G.,  1987, \mn@doi [\apj]
  {10.1086/164947}, \href
  {https://ui.adsabs.harvard.edu/abs/1987ApJ...313...42D} {313, 42}

\bibitem[\protect\citeauthoryear{{Etherington} et~al.,}{{Etherington}
  et~al.}{2022}]{No_lens_left_behind}
{Etherington} A.,  et~al., 2022, \mn@doi [\mnras] {10.1093/mnras/stac2639},
  \href {https://ui.adsabs.harvard.edu/abs/2022MNRAS.517.3275E} {517, 3275}

\bibitem[\protect\citeauthoryear{{Etherington} et~al.,}{{Etherington}
  et~al.}{2023}]{Beyond_bulge_halo}
{Etherington} A.,  et~al., 2023, \mn@doi [\mnras] {10.1093/mnras/stad582},
  \href {https://ui.adsabs.harvard.edu/abs/2023MNRAS.521.6005E} {521, 6005}

\bibitem[\protect\citeauthoryear{{Euclid Collaboration} et~al.,}{{Euclid
  Collaboration} et~al.}{2022}]{Euclid_Wide_Survey}
{Euclid Collaboration} et~al., 2022, \mn@doi [\aap]
  {10.1051/0004-6361/202141938}, \href
  {https://ui.adsabs.harvard.edu/abs/2022A&A...662A.112E} {662, A112}

\bibitem[\protect\citeauthoryear{{Euclid Collaboration} et~al.,}{{Euclid
  Collaboration} et~al.}{2024a}]{Euclid_Flagship_simulation}
{Euclid Collaboration} et~al., 2024a, \mn@doi [arXiv e-prints]
  {10.48550/arXiv.2405.13495}, \href
  {https://ui.adsabs.harvard.edu/abs/2024arXiv240513495E} {p. arXiv:2405.13495}

\bibitem[\protect\citeauthoryear{{Euclid Collaboration} et~al.,}{{Euclid
  Collaboration} et~al.}{2024b}]{EUCLID_CNN_lensfinder}
{Euclid Collaboration} et~al., 2024b, \mn@doi [\aap]
  {10.1051/0004-6361/202347244}, \href
  {https://ui.adsabs.harvard.edu/abs/2024A&A...681A..68E} {681, A68}

\bibitem[\protect\citeauthoryear{{Falco}, {Gorenstein}  \& {Shapiro}}{{Falco}
  et~al.}{1985}]{Mass_sheet_degeneracy}
{Falco} E.~E.,  {Gorenstein} M.~V.,   {Shapiro} I.~I.,  1985, \mn@doi [\apjl]
  {10.1086/184422}, \href
  {https://ui.adsabs.harvard.edu/abs/1985ApJ...289L...1F} {289, L1}

\bibitem[\protect\citeauthoryear{{Faure} et~al.,}{{Faure}
  et~al.}{2008}]{Faure_2008}
{Faure} C.,  et~al., 2008, \mn@doi [\apjs] {10.1086/526426}, \href
  {https://ui.adsabs.harvard.edu/abs/2008ApJS..176...19F} {176, 19}

\bibitem[\protect\citeauthoryear{{Ferrami} \& {Wyithe}}{{Ferrami} \&
  {Wyithe}}{2023}]{Ferrami_Lensing_bright_end}
{Ferrami} G.,  {Wyithe} J. S.~B.,  2023, \mn@doi [\mnras]
  {10.1093/mnrasl/slad050}, \href
  {https://ui.adsabs.harvard.edu/abs/2023MNRAS.523L..21F} {523, L21}

\bibitem[\protect\citeauthoryear{{Gavazzi}, {Treu}, {Rhodes}, {Koopmans},
  {Bolton}, {Burles}, {Massey}  \& {Moustakas}}{{Gavazzi}
  et~al.}{2007}]{Gavazzi_SLACS_2007}
{Gavazzi} R.,  {Treu} T.,  {Rhodes} J.~D.,  {Koopmans} L. V.~E.,  {Bolton}
  A.~S.,  {Burles} S.,  {Massey} R.~J.,   {Moustakas} L.~A.,  2007, \mn@doi
  [\apj] {10.1086/519237}, \href
  {https://ui.adsabs.harvard.edu/abs/2007ApJ...667..176G} {667, 176}

\bibitem[\protect\citeauthoryear{{Gavazzi}, {Treu}, {Koopmans}, {Bolton},
  {Moustakas}, {Burles}  \& {Marshall}}{{Gavazzi}
  et~al.}{2008}]{Jackpot_Gavazzi}
{Gavazzi} R.,  {Treu} T.,  {Koopmans} L. V.~E.,  {Bolton} A.~S.,  {Moustakas}
  L.~A.,  {Burles} S.,   {Marshall} P.~J.,  2008, \mn@doi [\apj]
  {10.1086/529541}, \href
  {https://ui.adsabs.harvard.edu/abs/2008ApJ...677.1046G} {677, 1046}

\bibitem[\protect\citeauthoryear{{Gavazzi}, {Treu}, {Marshall}, {Brault}  \&
  {Ruff}}{{Gavazzi} et~al.}{2012}]{Gavazzi_SL2S_2012}
{Gavazzi} R.,  {Treu} T.,  {Marshall} P.~J.,  {Brault} F.,   {Ruff} A.,  2012,
  \mn@doi [\apj] {10.1088/0004-637X/761/2/170}, \href
  {https://ui.adsabs.harvard.edu/abs/2012ApJ...761..170G} {761, 170}

\bibitem[\protect\citeauthoryear{{Gavazzi}, {Marshall}, {Treu}  \&
  {Sonnenfeld}}{{Gavazzi} et~al.}{2014}]{RINGFINDER_ALGORITHM}
{Gavazzi} R.,  {Marshall} P.~J.,  {Treu} T.,   {Sonnenfeld} A.,  2014, \mn@doi
  [\apj] {10.1088/0004-637X/785/2/144}, \href
  {https://ui.adsabs.harvard.edu/abs/2014ApJ...785..144G} {785, 144}

\bibitem[\protect\citeauthoryear{{Geng}, {Cao}, {Liu}, {Liu}, {Biesiada}  \&
  {Lian}}{{Geng} et~al.}{2021}]{Geng_2021}
{Geng} S.,  {Cao} S.,  {Liu} Y.,  {Liu} T.,  {Biesiada} M.,   {Lian} Y.,  2021,
  \mn@doi [\mnras] {10.1093/mnras/stab519}, \href
  {https://ui.adsabs.harvard.edu/abs/2021MNRAS.503.1319G} {503, 1319}

\bibitem[\protect\citeauthoryear{{Grillo}}{{Grillo}}{2010}]{Grillo_DM_fraction}
{Grillo} C.,  2010, \mn@doi [\apj] {10.1088/0004-637X/722/1/779}, \href
  {https://ui.adsabs.harvard.edu/abs/2010ApJ...722..779G} {722, 779}

\bibitem[\protect\citeauthoryear{{Grillo}, {Lombardi}  \& {Bertin}}{{Grillo}
  et~al.}{2008}]{Grillo_cosmo_param_lens_dyn}
{Grillo} C.,  {Lombardi} M.,   {Bertin} G.,  2008, \mn@doi [\aap]
  {10.1051/0004-6361:20077534}, \href
  {https://ui.adsabs.harvard.edu/abs/2008A&A...477..397G} {477, 397}

\bibitem[\protect\citeauthoryear{{Grillo}, {Gobat}, {Lombardi}  \&
  {Rosati}}{{Grillo} et~al.}{2009}]{Grillo_SLACS_2009}
{Grillo} C.,  {Gobat} R.,  {Lombardi} M.,   {Rosati} P.,  2009, \mn@doi [\aap]
  {10.1051/0004-6361/200811604}, \href
  {https://ui.adsabs.harvard.edu/abs/2009A&A...501..461G} {501, 461}

\bibitem[\protect\citeauthoryear{{Hartley}, {Flamary}, {Jackson}, {Tagore}  \&
  {Metcalf}}{{Hartley} et~al.}{2017}]{Support_Vector_Machines}
{Hartley} P.,  {Flamary} R.,  {Jackson} N.,  {Tagore} A.~S.,   {Metcalf} R.~B.,
   2017, \mn@doi [\mnras] {10.1093/mnras/stx1733}, \href
  {https://ui.adsabs.harvard.edu/abs/2017MNRAS.471.3378H} {471, 3378}

\bibitem[\protect\citeauthoryear{{He} et~al.,}{{He}
  et~al.}{2020}]{He_CNN_search}
{He} Z.,  et~al., 2020, \mn@doi [\mnras] {10.1093/mnras/staa1917}, \href
  {https://ui.adsabs.harvard.edu/abs/2020MNRAS.497..556H} {497, 556}

\bibitem[\protect\citeauthoryear{{Hogg}, {Blandford}, {Kundic}, {Fassnacht}  \&
  {Malhotra}}{{Hogg} et~al.}{1996}]{Hogg_lens_in_HST}
{Hogg} D.~W.,  {Blandford} R.,  {Kundic} T.,  {Fassnacht} C.~D.,   {Malhotra}
  S.,  1996, \mn@doi [\apjl] {10.1086/310213}, \href
  {https://ui.adsabs.harvard.edu/abs/1996ApJ...467L..73H} {467, L73}

\bibitem[\protect\citeauthoryear{{Holloway}, {Verma}, {Marshall}, {More}  \&
  {Tecza}}{{Holloway} et~al.}{2023}]{Holloway_2023}
{Holloway} P.,  {Verma} A.,  {Marshall} P.~J.,  {More} A.,   {Tecza} M.,  2023,
  \mn@doi [\mnras] {10.1093/mnras/stad2371}, \href
  {https://ui.adsabs.harvard.edu/abs/2023MNRAS.525.2341H} {525, 2341}

\bibitem[\protect\citeauthoryear{{Holloway}, {Marshall}, {Verma}, {More},
  {Ca{\~n}ameras}, {Jaelani}, {Ishida}  \& {Wong}}{{Holloway}
  et~al.}{2024}]{Holloway_ensemble_classifier}
{Holloway} P.,  {Marshall} P.~J.,  {Verma} A.,  {More} A.,  {Ca{\~n}ameras} R.,
   {Jaelani} A.~T.,  {Ishida} Y.,   {Wong} K.~C.,  2024, \mn@doi [\mnras]
  {10.1093/mnras/stae875}, \href
  {https://ui.adsabs.harvard.edu/abs/2024MNRAS.530.1297H} {530, 1297}

\bibitem[\protect\citeauthoryear{{Ivezic} et~al.,}{{Ivezic}
  et~al.}{2008}]{Vera_Rubin_LSST}
{Ivezic} Z.,  et~al., 2008, \mn@doi [Serbian Astronomical Journal]
  {10.2298/SAJ0876001I}, \href
  {https://ui.adsabs.harvard.edu/abs/2008SerAJ.176....1I} {176, 1}

\bibitem[\protect\citeauthoryear{{Ivezi{\'c}} et~al.,}{{Ivezi{\'c}}
  et~al.}{2019}]{LSST_Ivezic_2019}
{Ivezi{\'c}} {\v{Z}}.,  et~al., 2019, \mn@doi [\apj]
  {10.3847/1538-4357/ab042c}, \href
  {https://ui.adsabs.harvard.edu/abs/2019ApJ...873..111I} {873, 111}

\bibitem[\protect\citeauthoryear{{Jacobs}, {Glazebrook}, {Collett}, {More}  \&
  {McCarthy}}{{Jacobs} et~al.}{2017}]{Jacobs_CNN_CFHTLS}
{Jacobs} C.,  {Glazebrook} K.,  {Collett} T.,  {More} A.,   {McCarthy} C.,
  2017, \mn@doi [\mnras] {10.1093/mnras/stx1492}, \href
  {https://ui.adsabs.harvard.edu/abs/2017MNRAS.471..167J} {471, 167}

\bibitem[\protect\citeauthoryear{{Jacobs} et~al.,}{{Jacobs}
  et~al.}{2019a}]{Jacobs_catalog}
{Jacobs} C.,  et~al., 2019a, \mn@doi [\apjs] {10.3847/1538-4365/ab26b6}, \href
  {https://ui.adsabs.harvard.edu/abs/2019ApJS..243...17J} {243, 17}

\bibitem[\protect\citeauthoryear{{Jacobs} et~al.,}{{Jacobs}
  et~al.}{2019b}]{Jacobs_CNN_DES}
{Jacobs} C.,  et~al., 2019b, \mn@doi [\mnras] {10.1093/mnras/stz272}, \href
  {https://ui.adsabs.harvard.edu/abs/2019MNRAS.484.5330J} {484, 5330}

\bibitem[\protect\citeauthoryear{{Jacobs}, {Glazebrook}, {Qin}  \&
  {Collett}}{{Jacobs} et~al.}{2022}]{Jacobs_CNN_parameters_for_lensing}
{Jacobs} C.,  {Glazebrook} K.,  {Qin} A.~K.,   {Collett} T.,  2022, \mn@doi
  [Astronomy and Computing] {10.1016/j.ascom.2021.100535}, \href
  {https://ui.adsabs.harvard.edu/abs/2022A&C....3800535J} {38, 100535}

\bibitem[\protect\citeauthoryear{{Jaelani} et~al.,}{{Jaelani}
  et~al.}{2020}]{Jaelani_SuGOHI5}
{Jaelani} A.~T.,  et~al., 2020, \mn@doi [\mnras] {10.1093/mnras/staa1062},
  \href {https://ui.adsabs.harvard.edu/abs/2020MNRAS.495.1291J} {495, 1291}

\bibitem[\protect\citeauthoryear{{Jaelani}, {More}, {Wong}, {Inoue}, {Chao},
  {Premadi}  \& {Ca{\~n}ameras}}{{Jaelani} et~al.}{2023}]{Jaelani_SuGOHI10}
{Jaelani} A.~T.,  {More} A.,  {Wong} K.~C.,  {Inoue} K.~T.,  {Chao} D. C.~Y.,
  {Premadi} P.~W.,   {Ca{\~n}ameras} R.,  2023, \mn@doi [arXiv e-prints]
  {10.48550/arXiv.2312.07333}, \href
  {https://ui.adsabs.harvard.edu/abs/2023arXiv231207333J} {p. arXiv:2312.07333}

\bibitem[\protect\citeauthoryear{{Joseph} et~al.,}{{Joseph}
  et~al.}{2014}]{PCA_lens_search}
{Joseph} R.,  et~al., 2014, \mn@doi [\aap] {10.1051/0004-6361/201423365}, \href
  {https://ui.adsabs.harvard.edu/abs/2014A&A...566A..63J} {566, A63}

\bibitem[\protect\citeauthoryear{{Kochanek}}{{Kochanek}}{1996}]{Kochanek_1996}
{Kochanek} C.~S.,  1996, \mn@doi [\apj] {10.1086/177538}, \href
  {https://ui.adsabs.harvard.edu/abs/1996ApJ...466..638K} {466, 638}

\bibitem[\protect\citeauthoryear{{Koopmans} \& {Treu}}{{Koopmans} \&
  {Treu}}{2003}]{Koopmans_Treu_2003}
{Koopmans} L. V.~E.,  {Treu} T.,  2003, \mn@doi [\apj] {10.1086/345423}, \href
  {https://ui.adsabs.harvard.edu/abs/2003ApJ...583..606K} {583, 606}

\bibitem[\protect\citeauthoryear{{Koopmans} et~al.,}{{Koopmans}
  et~al.}{2009}]{Koopmans_bulge_halo_conspiracy}
{Koopmans} L.~V.~E.,  et~al., 2009, \mn@doi [\apjl]
  {10.1088/0004-637X/703/1/L51}, \href
  {https://ui.adsabs.harvard.edu/abs/2009ApJ...703L..51K} {703, L51}

\bibitem[\protect\citeauthoryear{{Kormann}, {Schneider}  \&
  {Bartelmann}}{{Kormann} et~al.}{1994}]{Kormann_Schneider_Bartelmann_1994}
{Kormann} R.,  {Schneider} P.,   {Bartelmann} M.,  1994, \aap, \href
  {https://ui.adsabs.harvard.edu/abs/1994A&A...284..285K} {284, 285}

\bibitem[\protect\citeauthoryear{{La Barbera}, {de Carvalho}, {de La Rosa},
  {Lopes}, {Kohl-Moreira}  \& {Capelato}}{{La Barbera}
  et~al.}{2010a}]{LaBarbera_SPIDER_I_Reff}
{La Barbera} F.,  {de Carvalho} R.~R.,  {de La Rosa} I.~G.,  {Lopes} P.~A.~A.,
  {Kohl-Moreira} J.~L.,   {Capelato} H.~V.,  2010a, \mn@doi [\mnras]
  {10.1111/j.1365-2966.2010.16850.x}, \href
  {https://ui.adsabs.harvard.edu/abs/2010MNRAS.408.1313L} {408, 1313}

\bibitem[\protect\citeauthoryear{{La Barbera}, {de Carvalho}, {de La Rosa}  \&
  {Lopes}}{{La Barbera} et~al.}{2010b}]{BarberaFPgrizYJHK}
{La Barbera} F.,  {de Carvalho} R.~R.,  {de La Rosa} I.~G.,   {Lopes} P.~A.~A.,
   2010b, \mn@doi [\mnras] {10.1111/j.1365-2966.2010.17091.x}, \href
  {https://ui.adsabs.harvard.edu/abs/2010MNRAS.408.1335L} {408, 1335}

\bibitem[\protect\citeauthoryear{{Lapi}, {Negrello}, {Gonz{\'a}lez-Nuevo},
  {Cai}, {De Zotti}  \& {Danese}}{{Lapi} et~al.}{2012}]{Lapi_2012}
{Lapi} A.,  {Negrello} M.,  {Gonz{\'a}lez-Nuevo} J.,  {Cai} Z.~Y.,  {De Zotti}
  G.,   {Danese} L.,  2012, \mn@doi [\apj] {10.1088/0004-637X/755/1/46}, \href
  {https://ui.adsabs.harvard.edu/abs/2012ApJ...755...46L} {755, 46}

\bibitem[\protect\citeauthoryear{{Liu}, {Mutch}, {Angel}, {Duffy}, {Geil},
  {Poole}, {Mesinger}  \& {Wyithe}}{{Liu} et~al.}{2016}]{Dragons_IV_UVLF}
{Liu} C.,  {Mutch} S.~J.,  {Angel} P.~W.,  {Duffy} A.~R.,  {Geil} P.~M.,
  {Poole} G.~B.,  {Mesinger} A.,   {Wyithe} J. S.~B.,  2016, \mn@doi [\mnras]
  {10.1093/mnras/stw1015}, \href
  {https://ui.adsabs.harvard.edu/abs/2016MNRAS.462..235L} {462, 235}

\bibitem[\protect\citeauthoryear{{Mao} \& {Schneider}}{{Mao} \&
  {Schneider}}{1998}]{Substructure_lens_galaxy}
{Mao} S.,  {Schneider} P.,  1998, \mn@doi [\mnras]
  {10.1046/j.1365-8711.1998.01319.x}, \href
  {https://ui.adsabs.harvard.edu/abs/1998MNRAS.295..587M} {295, 587}

\bibitem[\protect\citeauthoryear{{Marshall}, {Blandford}  \& {Sako}}{{Marshall}
  et~al.}{2005}]{Marshall_Blandford_Sako}
{Marshall} P.,  {Blandford} R.,   {Sako} M.,  2005, \mn@doi [\nar]
  {10.1016/j.newar.2005.08.009}, \href
  {https://ui.adsabs.harvard.edu/abs/2005NewAR..49..387M} {49, 387}

\bibitem[\protect\citeauthoryear{{Marshall}, {Hogg}, {Moustakas}, {Fassnacht},
  {Brada{\v{c}}}, {Schrabback}  \& {Blandford}}{{Marshall}
  et~al.}{2009}]{Marshall_model_detection}
{Marshall} P.~J.,  {Hogg} D.~W.,  {Moustakas} L.~A.,  {Fassnacht} C.~D.,
  {Brada{\v{c}}} M.,  {Schrabback} T.,   {Blandford} R.~D.,  2009, \mn@doi
  [\apj] {10.1088/0004-637X/694/2/924}, \href
  {https://ui.adsabs.harvard.edu/abs/2009ApJ...694..924M} {694, 924}

\bibitem[\protect\citeauthoryear{{Marshall} et~al.,}{{Marshall}
  et~al.}{2016}]{Marshall_citizien_science}
{Marshall} P.~J.,  et~al., 2016, \mn@doi [\mnras] {10.1093/mnras/stv2009},
  \href {https://ui.adsabs.harvard.edu/abs/2016MNRAS.455.1171M} {455, 1171}

\bibitem[\protect\citeauthoryear{{Mason} et~al.,}{{Mason}
  et~al.}{2015}]{Mason_2015}
{Mason} C.~A.,  et~al., 2015, \mn@doi [\apj] {10.1088/0004-637X/805/1/79},
  \href {https://ui.adsabs.harvard.edu/abs/2015ApJ...805...79M} {805, 79}

\bibitem[\protect\citeauthoryear{{Mondal}, {Saha}, {Windhorst}  \&
  {Jansen}}{{Mondal} et~al.}{2023}]{UV_cont_near_un}
{Mondal} C.,  {Saha} K.,  {Windhorst} R.~A.,   {Jansen} R.~A.,  2023, \mn@doi
  [\apj] {10.3847/1538-4357/acc110}, \href
  {https://ui.adsabs.harvard.edu/abs/2023ApJ...946...90M} {946, 90}

\bibitem[\protect\citeauthoryear{{More}, {Cabanac}, {More}, {Alard},
  {Limousin}, {Kneib}, {Gavazzi}  \& {Motta}}{{More}
  et~al.}{2012}]{ARCFINDER_More}
{More} A.,  {Cabanac} R.,  {More} S.,  {Alard} C.,  {Limousin} M.,  {Kneib}
  J.~P.,  {Gavazzi} R.,   {Motta} V.,  2012, \mn@doi [\apj]
  {10.1088/0004-637X/749/1/38}, \href
  {https://ui.adsabs.harvard.edu/abs/2012ApJ...749...38M} {749, 38}

\bibitem[\protect\citeauthoryear{{More} et~al.,}{{More}
  et~al.}{2016}]{More_citizien_science}
{More} A.,  et~al., 2016, \mn@doi [\mnras] {10.1093/mnras/stv1965}, \href
  {https://ui.adsabs.harvard.edu/abs/2016MNRAS.455.1191M} {455, 1191}

\bibitem[\protect\citeauthoryear{{Moustakas} et~al.,}{{Moustakas}
  et~al.}{2007}]{Moustakas_visual_inspection}
{Moustakas} L.~A.,  et~al., 2007, \mn@doi [\apjl] {10.1086/517930}, \href
  {https://ui.adsabs.harvard.edu/abs/2007ApJ...660L..31M} {660, L31}

\bibitem[\protect\citeauthoryear{{Nightingale} \& {Dye}}{{Nightingale} \&
  {Dye}}{2015}]{Nightingale_Dye_PyAutolens}
{Nightingale} J.~W.,  {Dye} S.,  2015, \mn@doi [\mnras]
  {10.1093/mnras/stv1455}, \href
  {https://ui.adsabs.harvard.edu/abs/2015MNRAS.452.2940N} {452, 2940}

\bibitem[\protect\citeauthoryear{{O'Donnell} et~al.,}{{O'Donnell}
  et~al.}{2022}]{DES_Strong_lens_odonnel}
{O'Donnell} J.~H.,  et~al., 2022, \mn@doi [\apjs] {10.3847/1538-4365/ac470b},
  \href {https://ui.adsabs.harvard.edu/abs/2022ApJS..259...27O} {259, 27}

\bibitem[\protect\citeauthoryear{{Oguri} \& {Marshall}}{{Oguri} \&
  {Marshall}}{2010}]{Oguri_Marshall}
{Oguri} M.,  {Marshall} P.~J.,  2010, \mn@doi [\mnras]
  {10.1111/j.1365-2966.2010.16639.x}, \href
  {https://ui.adsabs.harvard.edu/abs/2010MNRAS.405.2579O} {405, 2579}

\bibitem[\protect\citeauthoryear{{Oguri} et~al.,}{{Oguri}
  et~al.}{2012}]{Oguri_2012}
{Oguri} M.,  et~al., 2012, \mn@doi [\aj] {10.1088/0004-6256/143/5/120}, \href
  {https://ui.adsabs.harvard.edu/abs/2012AJ....143..120O} {143, 120}

\bibitem[\protect\citeauthoryear{{Oldham} \& {Auger}}{{Oldham} \&
  {Auger}}{2018}]{Oldham_DM_fraction}
{Oldham} L.~J.,  {Auger} M.~W.,  2018, \mn@doi [\mnras] {10.1093/mnras/sty065},
  \href {https://ui.adsabs.harvard.edu/abs/2018MNRAS.476..133O} {476, 133}

\bibitem[\protect\citeauthoryear{{Pawase}, {Courbin}, {Faure}, {Kokotanekova}
  \& {Meylan}}{{Pawase} et~al.}{2014}]{Pawase_visual_inspection}
{Pawase} R.~S.,  {Courbin} F.,  {Faure} C.,  {Kokotanekova} R.,   {Meylan} G.,
  2014, \mn@doi [\mnras] {10.1093/mnras/stu179}, \href
  {https://ui.adsabs.harvard.edu/abs/2014MNRAS.439.3392P} {439, 3392}

\bibitem[\protect\citeauthoryear{{Pei}}{{Pei}}{1995}]{1995_Pei}
{Pei} Y.~C.,  1995, \mn@doi [\apj] {10.1086/175290}, \href
  {https://ui.adsabs.harvard.edu/abs/1995ApJ...440..485P} {440, 485}

\bibitem[\protect\citeauthoryear{{Refsdal}}{{Refsdal}}{1964}]{Refsdal_1964}
{Refsdal} S.,  1964, \mn@doi [\mnras] {10.1093/mnras/128.4.307}, \href
  {https://ui.adsabs.harvard.edu/abs/1964MNRAS.128..307R} {128, 307}

\bibitem[\protect\citeauthoryear{{Sahu}, {Graham}  \& {Davis}}{{Sahu}
  et~al.}{2019}]{M_v_vel_disp_relation}
{Sahu} N.,  {Graham} A.~W.,   {Davis} B.~L.,  2019, \mn@doi [\apj]
  {10.3847/1538-4357/ab50b7}, \href
  {https://ui.adsabs.harvard.edu/abs/2019ApJ...887...10S} {887, 10}

\bibitem[\protect\citeauthoryear{{Schuldt}, {Chiriv{\`\i}}, {Suyu},
  {Y{\i}ld{\i}r{\i}m}, {Sonnenfeld}, {Halkola}  \& {Lewis}}{{Schuldt}
  et~al.}{2019}]{Schuldt_double_lens}
{Schuldt} S.,  {Chiriv{\`\i}} G.,  {Suyu} S.~H.,  {Y{\i}ld{\i}r{\i}m} A.,
  {Sonnenfeld} A.,  {Halkola} A.,   {Lewis} G.~F.,  2019, \mn@doi [\aap]
  {10.1051/0004-6361/201935042}, \href
  {https://ui.adsabs.harvard.edu/abs/2019A&A...631A..40S} {631, A40}

\bibitem[\protect\citeauthoryear{{Schwab}, {Bolton}  \& {Rappaport}}{{Schwab}
  et~al.}{2010}]{Schwab_Bolton_Rappaport}
{Schwab} J.,  {Bolton} A.~S.,   {Rappaport} S.~A.,  2010, \mn@doi [\apj]
  {10.1088/0004-637X/708/1/750}, \href
  {https://ui.adsabs.harvard.edu/abs/2010ApJ...708..750S} {708, 750}

\bibitem[\protect\citeauthoryear{{Scoville} et~al.,}{{Scoville}
  et~al.}{2007}]{HST_COSMOS}
{Scoville} N.,  et~al., 2007, \mn@doi [\apjs] {10.1086/516580}, \href
  {https://ui.adsabs.harvard.edu/abs/2007ApJS..172...38S} {172, 38}

\bibitem[\protect\citeauthoryear{{Serjeant}}{{Serjeant}}{2014}]{Serjeant_analytical_model}
{Serjeant} S.,  2014, \mn@doi [\apjl] {10.1088/2041-8205/793/1/L10}, \href
  {https://ui.adsabs.harvard.edu/abs/2014ApJ...793L..10S} {793, L10}

\bibitem[\protect\citeauthoryear{{Shajib}, {Treu}, {Birrer}  \&
  {Sonnenfeld}}{{Shajib} et~al.}{2021}]{Shajib_Dolphin}
{Shajib} A.~J.,  {Treu} T.,  {Birrer} S.,   {Sonnenfeld} A.,  2021, \mn@doi
  [\mnras] {10.1093/mnras/stab536}, \href
  {https://ui.adsabs.harvard.edu/abs/2021MNRAS.503.2380S} {503, 2380}

\bibitem[\protect\citeauthoryear{{Shajib} et~al.,}{{Shajib}
  et~al.}{2023}]{Shajib_lensing_H0}
{Shajib} A.~J.,  et~al., 2023, \mn@doi [\aap] {10.1051/0004-6361/202345878},
  \href {https://ui.adsabs.harvard.edu/abs/2023A&A...673A...9S} {673, A9}

\bibitem[\protect\citeauthoryear{{Shibuya}, {Ouchi}  \& {Harikane}}{{Shibuya}
  et~al.}{2015}]{Shibuya_L_size_rel}
{Shibuya} T.,  {Ouchi} M.,   {Harikane} Y.,  2015, \mn@doi [\apjs]
  {10.1088/0067-0049/219/2/15}, \href
  {https://ui.adsabs.harvard.edu/abs/2015ApJS..219...15S} {219, 15}

\bibitem[\protect\citeauthoryear{{Smith}, {Lucey}  \& {Conroy}}{{Smith}
  et~al.}{2015}]{SINFONI_survey}
{Smith} R.~J.,  {Lucey} J.~R.,   {Conroy} C.,  2015, \mn@doi [\mnras]
  {10.1093/mnras/stv518}, \href
  {https://ui.adsabs.harvard.edu/abs/2015MNRAS.449.3441S} {449, 3441}

\bibitem[\protect\citeauthoryear{{Sonnenfeld}}{{Sonnenfeld}}{2022a}]{Sonnenfeld_SL_Stat_III_inference_complete}
{Sonnenfeld} A.,  2022a, \mn@doi [\aap] {10.1051/0004-6361/202142301}, \href
  {https://ui.adsabs.harvard.edu/abs/2022A&A...659A.132S} {659, A132}

\bibitem[\protect\citeauthoryear{{Sonnenfeld}}{{Sonnenfeld}}{2022b}]{Sonnenfeld_SL_Stat_IV_inference_incomplete}
{Sonnenfeld} A.,  2022b, \mn@doi [\aap] {10.1051/0004-6361/202142467}, \href
  {https://ui.adsabs.harvard.edu/abs/2022A&A...659A.133S} {659, A133}

\bibitem[\protect\citeauthoryear{{Sonnenfeld} \& {Cautun}}{{Sonnenfeld} \&
  {Cautun}}{2021}]{Sonnenfeld_SL_Stat_I_innerstructure}
{Sonnenfeld} A.,  {Cautun} M.,  2021, \mn@doi [\aap]
  {10.1051/0004-6361/202140549}, \href
  {https://ui.adsabs.harvard.edu/abs/2021A&A...651A..18S} {651, A18}

\bibitem[\protect\citeauthoryear{{Sonnenfeld}, {Gavazzi}, {Suyu}, {Treu}  \&
  {Marshall}}{{Sonnenfeld} et~al.}{2013a}]{Sonnenfeld_SL2S_2013a}
{Sonnenfeld} A.,  {Gavazzi} R.,  {Suyu} S.~H.,  {Treu} T.,   {Marshall} P.~J.,
  2013a, \mn@doi [\apj] {10.1088/0004-637X/777/2/97}, \href
  {https://ui.adsabs.harvard.edu/abs/2013ApJ...777...97S} {777, 97}

\bibitem[\protect\citeauthoryear{{Sonnenfeld}, {Treu}, {Gavazzi}, {Suyu},
  {Marshall}, {Auger}  \& {Nipoti}}{{Sonnenfeld}
  et~al.}{2013b}]{Sonnenfeld_SL2S_2013}
{Sonnenfeld} A.,  {Treu} T.,  {Gavazzi} R.,  {Suyu} S.~H.,  {Marshall} P.~J.,
  {Auger} M.~W.,   {Nipoti} C.,  2013b, \mn@doi [\apj]
  {10.1088/0004-637X/777/2/98}, \href
  {https://ui.adsabs.harvard.edu/abs/2013ApJ...777...98S} {777, 98}

\bibitem[\protect\citeauthoryear{{Sonnenfeld}, {Treu}, {Marshall}, {Suyu},
  {Gavazzi}, {Auger}  \& {Nipoti}}{{Sonnenfeld}
  et~al.}{2015}]{Sonnenfeld_DM_fraction_SL2S}
{Sonnenfeld} A.,  {Treu} T.,  {Marshall} P.~J.,  {Suyu} S.~H.,  {Gavazzi} R.,
  {Auger} M.~W.,   {Nipoti} C.,  2015, \mn@doi [\apj]
  {10.1088/0004-637X/800/2/94}, \href
  {https://ui.adsabs.harvard.edu/abs/2015ApJ...800...94S} {800, 94}

\bibitem[\protect\citeauthoryear{{Sonnenfeld} et~al.,}{{Sonnenfeld}
  et~al.}{2018}]{Sonnenfeld_SuGOHI_I}
{Sonnenfeld} A.,  et~al., 2018, \mn@doi [\pasj] {10.1093/pasj/psx062}, \href
  {https://ui.adsabs.harvard.edu/abs/2018PASJ...70S..29S} {70, S29}

\bibitem[\protect\citeauthoryear{{Sonnenfeld}, {Jaelani}, {Chan}, {More},
  {Suyu}, {Wong}, {Oguri}  \& {Lee}}{{Sonnenfeld}
  et~al.}{2019}]{Sonnenfeld_IMF_SuGOHI}
{Sonnenfeld} A.,  {Jaelani} A.~T.,  {Chan} J.,  {More} A.,  {Suyu} S.~H.,
  {Wong} K.~C.,  {Oguri} M.,   {Lee} C.-H.,  2019, \mn@doi [\aap]
  {10.1051/0004-6361/201935743}, \href
  {https://ui.adsabs.harvard.edu/abs/2019A&A...630A..71S} {630, A71}

\bibitem[\protect\citeauthoryear{{Sonnenfeld} et~al.,}{{Sonnenfeld}
  et~al.}{2020}]{Sonnenfeld_citizien_science}
{Sonnenfeld} A.,  et~al., 2020, \mn@doi [\aap] {10.1051/0004-6361/202038067},
  \href {https://ui.adsabs.harvard.edu/abs/2020A&A...642A.148S} {642, A148}

\bibitem[\protect\citeauthoryear{{Sonnenfeld}, {Li}, {Despali}, {Gavazzi},
  {Shajib}  \& {Taylor}}{{Sonnenfeld} et~al.}{2023}]{Sonnenfeld_SL_selection}
{Sonnenfeld} A.,  {Li} S.-S.,  {Despali} G.,  {Gavazzi} R.,  {Shajib} A.~J.,
  {Taylor} E.~N.,  2023, \mn@doi [\aap] {10.1051/0004-6361/202346026}, \href
  {https://ui.adsabs.harvard.edu/abs/2023A&A...678A...4S} {678, A4}

\bibitem[\protect\citeauthoryear{{Spergel} et~al.,}{{Spergel}
  et~al.}{2015}]{Roman_WFIRST}
{Spergel} D.,  et~al., 2015, \mn@doi [arXiv e-prints]
  {10.48550/arXiv.1503.03757}, \href
  {https://ui.adsabs.harvard.edu/abs/2015arXiv150303757S} {p. arXiv:1503.03757}

\bibitem[\protect\citeauthoryear{{Stein}, {Blaum}, {Harrington}, {Medan}  \&
  {Luki{\'c}}}{{Stein} et~al.}{2022}]{CNN_unspervised_search}
{Stein} G.,  {Blaum} J.,  {Harrington} P.,  {Medan} T.,   {Luki{\'c}} Z.,
  2022, \mn@doi [\apj] {10.3847/1538-4357/ac6d63}, \href
  {https://ui.adsabs.harvard.edu/abs/2022ApJ...932..107S} {932, 107}

\bibitem[\protect\citeauthoryear{Stockmann et~al.,}{Stockmann
  et~al.}{2021}]{Stockmann_2021}
Stockmann M.,  et~al., 2021, \mn@doi [The Astrophysical Journal]
  {10.3847/1538-4357/abce66}, 908, 135

\bibitem[\protect\citeauthoryear{{Suyu} et~al.,}{{Suyu}
  et~al.}{2017}]{Suyu_COSMOGRAIL}
{Suyu} S.~H.,  et~al., 2017, \mn@doi [\mnras] {10.1093/mnras/stx483}, \href
  {https://ui.adsabs.harvard.edu/abs/2017MNRAS.468.2590S} {468, 2590}

\bibitem[\protect\citeauthoryear{{Tan} et~al.,}{{Tan}
  et~al.}{2024}]{Project_Dinos}
{Tan} C.~Y.,  et~al., 2024, \mn@doi [\mnras] {10.1093/mnras/stae884}, \href
  {https://ui.adsabs.harvard.edu/abs/2024MNRAS.530.1474T} {530, 1474}

\bibitem[\protect\citeauthoryear{{Tanaka} et~al.,}{{Tanaka}
  et~al.}{2016}]{Tanaka_double_lens}
{Tanaka} M.,  et~al., 2016, \mn@doi [\apjl] {10.3847/2041-8205/826/2/L19},
  \href {https://ui.adsabs.harvard.edu/abs/2016ApJ...826L..19T} {826, L19}

\bibitem[\protect\citeauthoryear{{Thuruthipilly}, {Zadrozny}, {Pollo}  \&
  {Biesiada}}{{Thuruthipilly} et~al.}{2022}]{Self_attention_ML}
{Thuruthipilly} H.,  {Zadrozny} A.,  {Pollo} A.,   {Biesiada} M.,  2022,
  \mn@doi [\aap] {10.1051/0004-6361/202142463}, \href
  {https://ui.adsabs.harvard.edu/abs/2022A&A...664A...4T} {664, A4}

\bibitem[\protect\citeauthoryear{{Tran} et~al.,}{{Tran}
  et~al.}{2022}]{AGEL_Survey}
{Tran} K.-V.~H.,  et~al., 2022, \mn@doi [\aj] {10.3847/1538-3881/ac7da2}, \href
  {https://ui.adsabs.harvard.edu/abs/2022AJ....164..148T} {164, 148}

\bibitem[\protect\citeauthoryear{{Trenti} \& {Stiavelli}}{{Trenti} \&
  {Stiavelli}}{2008}]{TrentiCosmic_Variance}
{Trenti} M.,  {Stiavelli} M.,  2008, \mn@doi [\apj] {10.1086/528674}, \href
  {https://ui.adsabs.harvard.edu/abs/2008ApJ...676..767T} {676, 767}

\bibitem[\protect\citeauthoryear{{Treu}}{{Treu}}{2010}]{Treu_review_2010}
{Treu} T.,  2010, \mn@doi [\araa] {10.1146/annurev-astro-081309-130924}, \href
  {https://ui.adsabs.harvard.edu/abs/2010ARA&A..48...87T} {48, 87}

\bibitem[\protect\citeauthoryear{{Treu} \& {Koopmans}}{{Treu} \&
  {Koopmans}}{2002}]{Treu_Koopmans_2002}
{Treu} T.,  {Koopmans} L. V.~E.,  2002, \mn@doi [\apj] {10.1086/341216}, \href
  {https://ui.adsabs.harvard.edu/abs/2002ApJ...575...87T} {575, 87}

\bibitem[\protect\citeauthoryear{{Treu}, {Stiavelli}, {Bertin}, {Casertano}  \&
  {M{\o}ller}}{{Treu} et~al.}{2001}]{2001TreuFP}
{Treu} T.,  {Stiavelli} M.,  {Bertin} G.,  {Casertano} S.,   {M{\o}ller} P.,
  2001, \mn@doi [\mnras] {10.1046/j.1365-8711.2001.04720.x}, \href
  {https://ui.adsabs.harvard.edu/abs/2001MNRAS.326..237T} {326, 237}

\bibitem[\protect\citeauthoryear{{Treu}, {Auger}, {Koopmans}, {Gavazzi},
  {Marshall}  \& {Bolton}}{{Treu} et~al.}{2010}]{Treu_IMF}
{Treu} T.,  {Auger} M.~W.,  {Koopmans} L. V.~E.,  {Gavazzi} R.,  {Marshall}
  P.~J.,   {Bolton} A.~S.,  2010, \mn@doi [\apj]
  {10.1088/0004-637X/709/2/1195}, \href
  {https://ui.adsabs.harvard.edu/abs/2010ApJ...709.1195T} {709, 1195}

\bibitem[\protect\citeauthoryear{{Tu} et~al.,}{{Tu}
  et~al.}{2009}]{Tu_double_lens}
{Tu} H.,  et~al., 2009, \mn@doi [\aap] {10.1051/0004-6361/200911963}, \href
  {https://ui.adsabs.harvard.edu/abs/2009A&A...501..475T} {501, 475}

\bibitem[\protect\citeauthoryear{{Turner}, {Ostriker}  \& {Gott}}{{Turner}
  et~al.}{1984}]{Turner_1984}
{Turner} E.~L.,  {Ostriker} J.~P.,   {Gott} J.~R. I.,  1984, \mn@doi [\apj]
  {10.1086/162379}, \href
  {https://ui.adsabs.harvard.edu/abs/1984ApJ...284....1T} {284, 1}

\bibitem[\protect\citeauthoryear{{Vegetti}, {Koopmans}, {Bolton}, {Treu}  \&
  {Gavazzi}}{{Vegetti} et~al.}{2010}]{Vegetti_Substructure_Lensing}
{Vegetti} S.,  {Koopmans} L.~V.~E.,  {Bolton} A.,  {Treu} T.,   {Gavazzi} R.,
  2010, \mn@doi [\mnras] {10.1111/j.1365-2966.2010.16865.x}, \href
  {https://ui.adsabs.harvard.edu/abs/2010MNRAS.408.1969V} {408, 1969}

\bibitem[\protect\citeauthoryear{{Walsh}, {Carswell}  \& {Weymann}}{{Walsh}
  et~al.}{1979}]{Walsh_first_observed_lens}
{Walsh} D.,  {Carswell} R.~F.,   {Weymann} R.~J.,  1979, \mn@doi [\nat]
  {10.1038/279381a0}, \href
  {https://ui.adsabs.harvard.edu/abs/1979Natur.279..381W} {279, 381}

\bibitem[\protect\citeauthoryear{{Webster}}{{Webster}}{1991}]{Webster_AGN_LF}
{Webster} R.,  1991, in {Crampton} D.,  ed.,  Astronomical Society of the
  Pacific Conference Series Vol. 21, The Space Distribution of Quasars. p.~160

\bibitem[\protect\citeauthoryear{{Windhorst} et~al.,}{{Windhorst}
  et~al.}{2023}]{Windhorst_2022_PEARLS}
{Windhorst} R.~A.,  et~al., 2023, \mn@doi [\aj] {10.3847/1538-3881/aca163},
  \href {https://ui.adsabs.harvard.edu/abs/2023AJ....165...13W} {165, 13}

\bibitem[\protect\citeauthoryear{{Wong} et~al.,}{{Wong}
  et~al.}{2018}]{Wong_SuGOHI2}
{Wong} K.~C.,  et~al., 2018, \mn@doi [\apj] {10.3847/1538-4357/aae381}, \href
  {https://ui.adsabs.harvard.edu/abs/2018ApJ...867..107W} {867, 107}

\bibitem[\protect\citeauthoryear{{Wong} et~al.,}{{Wong}
  et~al.}{2020}]{H0LiCOW_Wong}
{Wong} K.~C.,  et~al., 2020, \mn@doi [\mnras] {10.1093/mnras/stz3094}, \href
  {https://ui.adsabs.harvard.edu/abs/2020MNRAS.498.1420W} {498, 1420}

\bibitem[\protect\citeauthoryear{{Wong}, {Chan}, {Chao}, {Jaelani}, {Kayo},
  {Lee}, {More}  \& {Oguri}}{{Wong} et~al.}{2022}]{Wong_SUGOHI}
{Wong} K.~C.,  {Chan} J. H.~H.,  {Chao} D. C.~Y.,  {Jaelani} A.~T.,  {Kayo} I.,
   {Lee} C.-H.,  {More} A.,   {Oguri} M.,  2022, \mn@doi [\pasj]
  {10.1093/pasj/psac065}, \href
  {https://ui.adsabs.harvard.edu/abs/2022PASJ...74.1209W} {74, 1209}

\bibitem[\protect\citeauthoryear{{Wyithe} \& {Loeb}}{{Wyithe} \&
  {Loeb}}{2002a}]{Wyithe_Loeb_AGN_LF}
{Wyithe} J. S.~B.,  {Loeb} A.,  2002a, \mn@doi [\nat] {10.1038/nature00794},
  \href {https://ui.adsabs.harvard.edu/abs/2002Natur.417..923W} {417, 923}

\bibitem[\protect\citeauthoryear{{Wyithe} \& {Loeb}}{{Wyithe} \&
  {Loeb}}{2002b}]{Wyithe_SDSS_2002}
{Wyithe} J. S.~B.,  {Loeb} A.,  2002b, \mn@doi [\apj] {10.1086/342181}, \href
  {https://ui.adsabs.harvard.edu/abs/2002ApJ...577...57W} {577, 57}

\bibitem[\protect\citeauthoryear{{Wyithe}, {Turner}  \& {Spergel}}{{Wyithe}
  et~al.}{2001}]{Wyithe_gNFW}
{Wyithe} J.~S.~B.,  {Turner} E.~L.,   {Spergel} D.~N.,  2001, \mn@doi [\apj]
  {10.1086/321437}, \href
  {https://ui.adsabs.harvard.edu/abs/2001ApJ...555..504W} {555, 504}

\bibitem[\protect\citeauthoryear{{Wyithe}, {Yan}, {Windhorst}  \&
  {Mao}}{{Wyithe} et~al.}{2011}]{Wyithe_Nature_2011}
{Wyithe} J. S.~B.,  {Yan} H.,  {Windhorst} R.~A.,   {Mao} S.,  2011, \mn@doi
  [\nat] {10.1038/nature09619}, \href
  {https://ui.adsabs.harvard.edu/abs/2011Natur.469..181W} {469, 181}

\bibitem[\protect\citeauthoryear{{Zwicky}}{{Zwicky}}{1937}]{Zwicky_lenses}
{Zwicky} F.,  1937, \mn@doi [Physical Review] {10.1103/PhysRev.51.290}, \href
  {https://ui.adsabs.harvard.edu/abs/1937PhRv...51..290Z} {51, 290}

\bibitem[\protect\citeauthoryear{{de Vaucouleurs}}{{de
  Vaucouleurs}}{1948}]{DeVaucouleurs1948}
{de Vaucouleurs} G.,  1948, Annales d'Astrophysique, \href
  {https://ui.adsabs.harvard.edu/abs/1948AnAp...11..247D} {11, 247}

\bibitem[\protect\citeauthoryear{{de Vaucouleurs}}{{de
  Vaucouleurs}}{1953}]{DeVaucouleurs1953}
{de Vaucouleurs} G.,  1953, \mn@doi [\mnras] {10.1093/mnras/113.2.134}, \href
  {https://ui.adsabs.harvard.edu/abs/1953MNRAS.113..134D} {113, 134}

\bibitem[\protect\citeauthoryear{{van der Wel} et~al.,}{{van der Wel}
  et~al.}{2014}]{van_der_Wel_2014}
{van der Wel} A.,  et~al., 2014, \mn@doi [\apjl] {10.1088/2041-8205/792/1/L6},
  \href {https://ui.adsabs.harvard.edu/abs/2014ApJ...792L...6V} {792, L6}

\makeatother
\end{thebibliography}

\appendix

\section{Fitting the ML-based samples}
\begin{figure}
  \includegraphics[width=\linewidth]{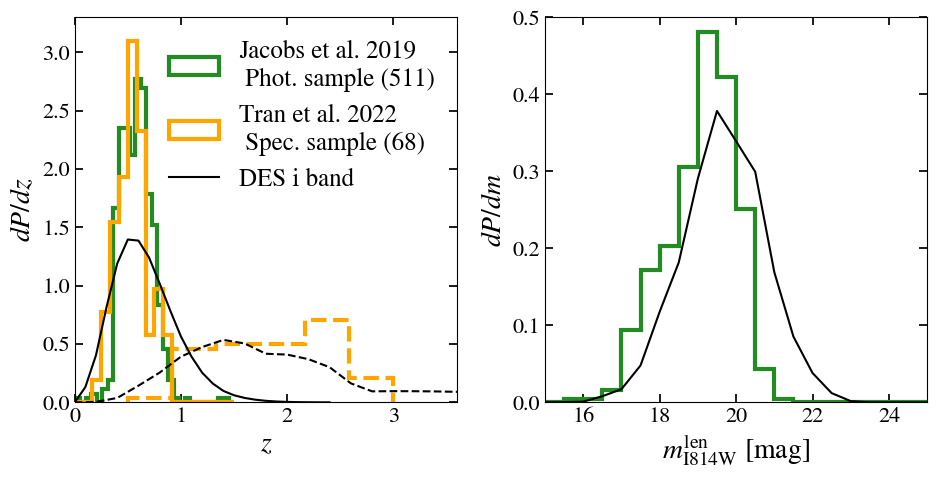}
    \caption{
    \REPLY{
    Fitting the observed DES lens distribution obtained from ML based search by 
    \citealt{Jacobs_catalog} with spectroscopic follow up in \citealt{AGEL_Survey}.
    To fit it we had to impose a VDF redshift evolution of $\dv{\sigma_\star}{z} = 1$ and $\dv{\alpha}{z} = 1$.
    }}
    \label{fig:ML_fit_VDF}
\end{figure}

\REPLY{
In this appendix, we explore a choice of VDF parameters that fit the observed DES lens distribution obtained from ML based search by \citealt{Jacobs_catalog}, with spectroscopic follow up in \citealt{AGEL_Survey}.
We find that a redshift evolution of the VDF parameters $\sigma_\star$ and $\alpha$ linear in $z$ is compatible with the data, as shown in Figure \ref{fig:ML_fit_VDF}.
Such evolution of the VDF is hard to justify physically, given for example the negative redshift evolution of the LF slope and a slowly varying luminosity-velocity dispersion relation (e.g., see \citealt{Bouwens21_data}).
This gives more weight to the possibility that this ML search was biased towards higher redshift lenses.
}

\section{Comparison with Collett 2015 and Euclid Collaboration forecasts}
\begin{figure*}
  \includegraphics[width=\linewidth]{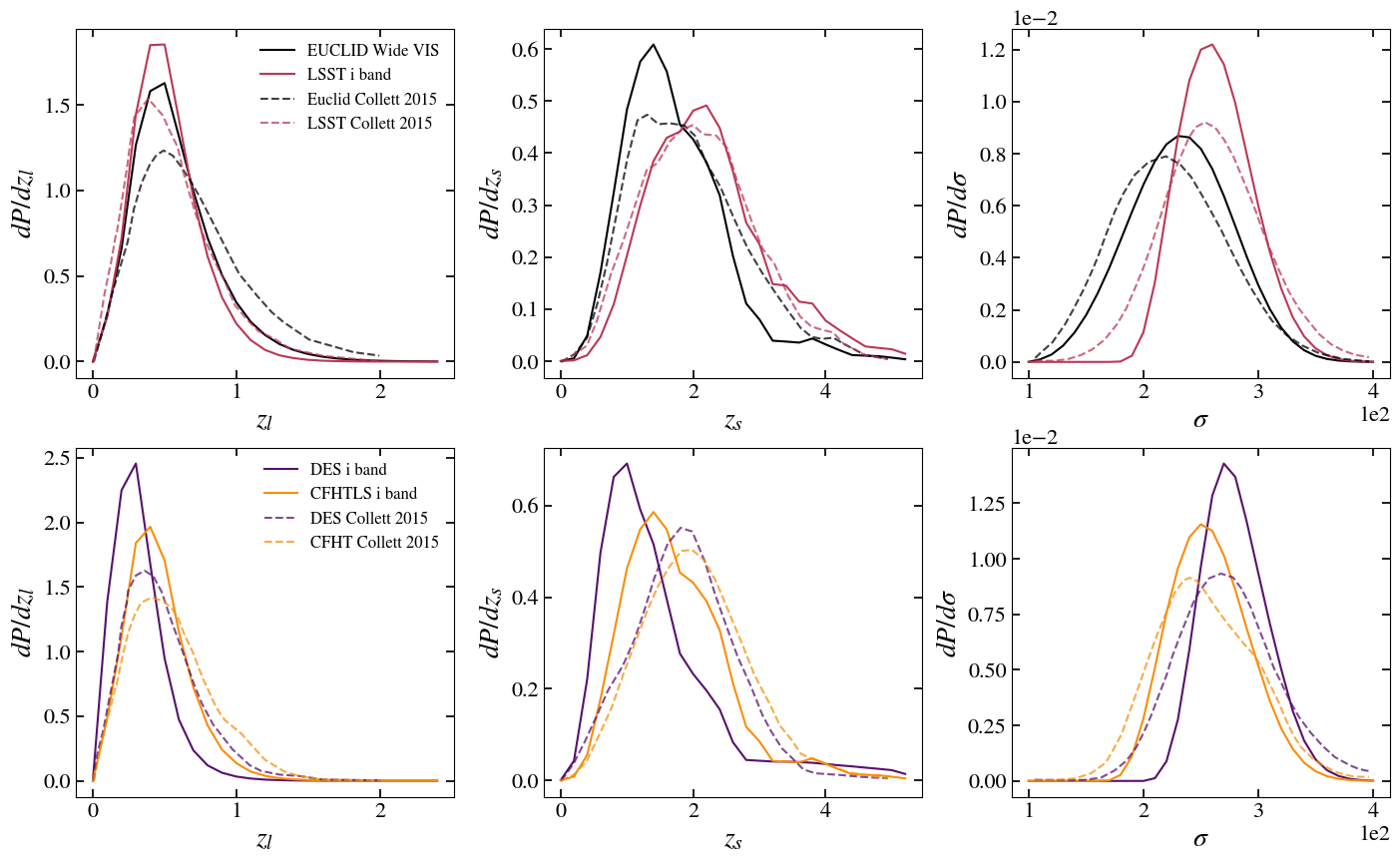}
    \caption{
    \REPLY{
    Comparison between the lens and source distributions predicted by our model (solid) and the results in \citealt{Collett_2015} (dashed).
    The left and middle panels show the redshift distributions of the lens and source populations, respectively. 
    The right panel show the velocity dispersion. 
    Black shows Euclid, red shows Vera Rubin LSST (upper), purple shows DES and orange shows the CFHTLS field (lower).
    }}
    \label{fig:comp_COLLETT}
\end{figure*}

\REPLY
{
In this appendix, we compare the lens and source distributions predicted by our model with the results from \citealt{Collett_2015}.
Figure \ref{fig:comp_COLLETT} shows that the relative change between the expected distributions in two given surveys is coherent between the two models.
Nonetheless, \citealt{Collett_2015} show on average a higher density of high redshift sources and a small excess of low mass lenses. This can probably be mainly reduced to the differences in the lens and source mass and densities assumed in the two models, even though a detailed comparison would require running several tests on the pipeline of \citealt{Collett_2015}.
The forecasts for the Euclid mission strong lens yield are obtained from \citealt{Collett_2015} 
(\citealt{Euclid_Wide_Survey}, \citealt{Euclid_Programme_2024}). 
Furthermore, we can compare our model results to the Euclid collaboration fiducial catalogue of mock galaxy scale strong lenses described in \citealt{EUCLID_CNN_lensfinder}, led by Leuzzi.
This catalogue used to train the neural network to recognize galaxy-galaxy lenses, is built combining lenses drawn from the Flagship simulation (\citealt{Euclid_Flagship_simulation}, led by Castander), and sources drawn from the Hubble Ultra Deep Field (UDF; \citealt{Hubble_UDF}), and the lensed image(s) are selected imposing $\mu_\text{arc}=1.6$ and a minimum source angular size of 20 pixels.
The resulting lens redshift and Einstein radius distributions are consistent with our model predictions for the Euclid VIS band, while they present an excess of high redshift sources above $z_s>2$ (cfr. Fig. \ref{fig:EUCLID_stat} with Fig. 1 in \citealt{EUCLID_CNN_lensfinder})
}

\bsp  
\label{lastpage}
\end{document}